\documentclass[fleqn,usenatbib]{mnras}
\pdfoutput=1
\usepackage[T1]{fontenc}
\DeclareRobustCommand{\VAN}[3]{#2}
\let\VANthebibliography\thebibliography
\def\thebibliography{\DeclareRobustCommand{\VAN}[3]{##3}\VANthebibliography}

\usepackage{graphicx,url}	

\usepackage{newtxtext,newtxmath}
\usepackage{tikz,xcolor,hyperref}

\definecolor{lime}{HTML}{A6CE39}
\DeclareRobustCommand{\orcidicon}{%
	\begin{tikzpicture}
	\draw[lime, fill=lime] (0,0) 
	circle [radius=0.16] 
	node[white] {{\fontfamily{qag}\selectfont \tiny ID}};
	\draw[white, fill=white] (-0.0625,0.095) 
	circle [radius=0.007];
	\end{tikzpicture}
	\hspace{-2mm}
}

\foreach \x in {A, ..., Z}{%
	\expandafter\xdef\csname orcid\x\endcsname{\noexpand\href{https://orcid.org/\csname orcidauthor\x\endcsname}{\noexpand\orcidicon}}
}

\title[Particle acceleration by dispersive Alfven waves]
{Particle acceleration by sub-{ proton} cyclotron frequency spectrum of dispersive Alfven waves in inhomogeneous solar coronal plasmas}
\author[D. Tsiklauri]{D. Tsiklauri,\thanks{E-mail: d.tsiklauri@qmul.ac.uk}\orcidA{}
\\
Department of Physics and Astronomy, Queen Mary University of London, Mile End Road, London E1 4NS, United Kingdom}
\date{Accepted 2023 December 21. Received 2023 November 26; in original form 2023 September 03}
\pubyear{2024}

\begin{document}
\label{firstpage}
\pagerange{\pageref{firstpage}--\pageref{lastpage}}
\maketitle

\begin{abstract}
The problem  of explaining observed soft X-ray fluxes during solar flares, 
which invokes acceleration of large fraction of electrons, if the acceleration takes places at the solar coronal loop-top,  can potentially be solved by postulating that flare at loop-top creates dispersive Alfven waves (DAWs) which propagate towards the foot-points. As DAWs move in progressively denser parts of the loop (due to gravitational stratification) the large fraction of electrons is no longer needed. Here we extend our previous results by considering $f ^{-1}$ frequency spectrum of DAWs and add ${\rm He^{++}}$ ions using fully kinetic particle-in-cell (PIC) simulations. We consider cases when transverse density gradient is  in the range ${ 4-40} c/\omega_{\rm { pe}}$ and DAW driving frequency is $0.3-0.6\omega_{\rm { cp}}$. We find that (i) The frequency spectrum case does not  affect electron acceleration fraction in the like-to-like cases, but few times larger percentage of ${\rm He^{++}}$ heating is seen due to ion cyclotron resonance; (ii) In cases when counter propagating DAWs collide multiple-times, much larger electron and ion acceleration fractions are found, but the process is intermittent in time. This is because intensive heating (temperature increase) makes the-above-thermal-fraction smaller; Also more isotropic velocity distributions are seen; (iii) Development of kink oscillations occurs when DAWs collide; (iv) Scaling of the magnetic fluctuations power spectrum steepening in the higher-density regions is seen, due to wave refraction. 
{ Our PIC runs produce much steeper slopes than the orginal spectrum, indicating that the} electron-scale physics has a notable effect of DAW spectrum evolution.
\end{abstract}




\section{Introduction}

Alfven waves play an important role in solar atmosphere as a means of transporting energy from lower to higher layers 
\citep{1998A&A...338.1118R}.
On large spatial scales, these waves may contribute to solving solar coronal heating problem 
in magnetohydrodynamic (MHD) approximation 
\citep{1983A&A...117..220H,2022MNRAS.510.1910B,2022MNRAS.510.2618B}.
On small spatial scales, i.e. in the kinetic approximation such waves
may play a major role in resolving problems arising in
quantifying observed energetic, super-thermal particles \citep{2005A&A...435.1105T,Fletcher_2008,2009ApJ...693.1494M}.

Some of relevant works on kinetic Alfven waves include the following:
Starting with books \citet{wu2012} and \citet{wu2020} discuss
 Kinetic Alfven wave aspects such as theory, experiments, observations and practical applications.
A review of Earth magnetospheric aspects of kinetic Alfven waves is presented in
\citet{2000SSRv...92..423S}.  Some recent works include:
\citet{PhysRevLett.128.025101} have studied kinetic-scale heating by Alfven Waves in magnetic shears. With first-principles kinetic simulations, they show that a large-scale Alfven wave (AW) propagating in an inhomogeneous background decays into kinetic Alfven waves (KAWs), triggering ion and electron energization. They  demonstrated that the two species can access unequal amounts of the initial AW energy, experiencing differential heating. 
\citet{maiorano_settino_malara_pezzi_pucci_valentini_2020} study
Kinetic Alfven wave generation 
by velocity shear in collisionless plasmas.
The evolution of a linearly polarized, 
long-wavelength Alfven wave propagating in a collisionless magnetized plasma with a sheared parallel-directed velocity flow has been studied by means of two-dimensional hybrid Vlasov-Maxwell (HVM) simulations.
By analyzing both the polarization and group velocity of perturbations in the shear regions, they identify them as KAWs.
In the moderate amplitude run, kinetic effects distort the proton distribution function in the shear region. This leads to the formation of a proton beam, at the Alfven speed and parallel to the magnetic field.
\citet{Xiang_2019} studied resonant mode conversion of AWs into KAWs when there is an arbitrary angle between background magnetic field and plasma 
density gradient in plasma with two different temperature species. 
\citet{Xiang_2019} find that the efficiency of mode conversion 
depends on the angle, and spatial scales such as 
the density inhomogeneous gradient scale, inverse of parallel wavenumber, and  electron/ion temperature ratio. 
\citet{Malara_2019} studied electron heating by KAWs in solar coronal loop turbulence. A test-particle model describing the energization of electrons in a turbulent plasma was presented. A fluctuating electric field component parallel to the background magnetic field, with properties similar to those of KAWs, is assumed to be present at scales of the order of the proton Larmor radius. Electrons were found to be stochastically accelerated by multiple interactions with such fluctuations, reaching energies of the order of $10^2$ eV within tens to hundreds of seconds, depending on the turbulence amplitude.
\citet{PhysRevE.96.023201} considered Turbulence generation during the head-on collision of Alfvenic wave packets in 
the context of Moffatt and Parker problem.
\citet{pezzi_2017_1}  have extended the work in the same context by including Hall magnetohydrodynamics and two hybrid kinetic Vlasov-Maxwell numerical models.
\citet{Pezzi_2017} considered the interaction of two colliding Alfven wave packets and found that the extension to include compressive and kinetic effects, while maintaining the gross characteristics of the simpler classic formulation, also reveals interesting features beyond MHD treatment. 
\citet{Valentini2017} studied a transition to kinetic turbulence at proton scales driven by large-amplitude kinetic Alfven fluctuations. 
 The transition from quasi-linear to turbulent regimes was investigated, focusing in particular on the development of important non-Maxwellian features in the proton distribution function driven by KAW fluctuations.
\citet{Pucci2016}
studied a model where an initial Alfven wave propagates inside an equilibrium structure which is inhomogeneous in the direction perpendicular to the equilibrium magnetic field. In a previous paper this situation has been considered in a particular configuration where the initial wave vector is parallel to the magnetic field and the wave is polarized perpendicular to the inhomogeneity direction. In \citet{Pucci2016} 
they consider other configurations, with a different polarization and possible initial oblique propagation.
\citet{Vasconez_2015} 
considered kinetic Alfven wave generation by large-scale phase-mixing and
found that kinetic simulations show that KAWs modify the ion distribution function, generating temperature anisotropy of both parallel and perpendicular to the local magnetic field as well as particle beams aligned along the local magnetic field. 
\citet{2003PhPl...10.3787W,2004JGRA..109.6211W} investigated the magnetic field-aligned acceleration of auroral super-thermal electrons by KAWs. 
In the solar physics context,
\citet{1999ApJ...511..958W} studied the nonuniform heating of magnetized plasmas in the solar atmosphere with KAWs using
the drift kinetic equation. They investigated electron heating mechanism
and proposed that it can in principle explain X-ray brightness distributions
observed in solar coronal loops.
\citet{2003ApJ...596..656W} investigated 
dissipation of KAWs in a solar polar plume placed in so-called coronal hole. 
They found that  electron heating produced by KAW dissipation can
in principle balance radiative losses of the plume. 
\citet{2007ApJ...659L.181W} tackled the problem
of explaining the observations that
sunspots have a higher temperature than the ambient quiet Sun in the upper chromosphere, although they appear dark in the photosphere.
\citet{2007ApJ...659L.181W} showed  above 850 km, in the upper chromosphere, KAW dissipation is the main heating mechanism at such heights.
Observations of the solar corona find that the heavy ion species experience anisotropic heating that is primarily across the magnetic field.
\citet{2006A&A...452L...7W,2007ApJ...659.1693W} investigated the nonlinear interaction of heavy ions with KAWs.
Using three-component plasma model with electrons, protons, and heavy ions.
Results show that heavy ions are energized across the magnetic field.
\citet{1996PhRvL..77.4346W,1997PhPl....4..611W} discuss a
new type of density soliton that is discovered in data from the Freja satellite.
A good case is made for it to be KAW based explanation.

Another interesting feature related to KAWs if found in Earth's outer radiation belt. The term time domain structures (TDSs) refers to packets of 
greater than 1-ms duration intense electric field spikes detected by Van Allen Probes in the Earth's outer radiation belt.
Phase space holes, double layers and other solitary electric field structures, referred to as time domain structures (TDSs), often occur around dipolarization fronts in the Earth's inner magnetosphere. 
In \citet{https://doi.org/10.1029/2020JA028643}, authors demonstrate that TDSs can be excited by electrons in nonlinear Landau resonance with kinetic Alfven waves.
\citet{Hughes_2017} considered kinetic Alfven turbulence and electron and ion heating by PIC simulations. They find that in contrast to dissipation by whistler turbulence, the maximum ion heating rate due to kinetic Alfven turbulence is substantially greater than the maximum electron heating rate. It is also found that the Landau wave-particle resonance is a likely heating mechanism for the electrons and may also contribute to ion heating.

There are also studies that examine causal connection between KAWs and
magnetic reconnection is 2D \citep{https://doi.org/10.1029/2017JA025071} 
and 3D \citep{https://doi.org/10.1002/2016JA022505}.
Specific effects related to KAW eigenmode in 
magnetosphere magnetic reconnection has been studied by \citet{Dai2022}.
\citet{https://doi.org/10.1002/2016GL071044} provided KAW explanation of the 
Hall fields in magnetic reconnection.

Coming back to solar flares, the observed soft X-ray flux during  flares is produced by electron bremsstrahlung, when accelerated electrons that move from magnetic loop top to the foot-points are slowed
 down by dense layers of the sun. In order to explain the observed soft X-ray flux during solar flares, if electron acceleration happens at loop top, nearly 100\% electrons need to be accelerated \citep{2020Sci...367..278F,2022Natur.606..674F}. No acceleration mechanism is known with such high efficiency. This problem can potentially be solved by postulating \citep{Fletcher_2008} that flare at loop top creates dispersive Alfven waves (DAWs) which then propagate towards the foot-points, and as they move in progressively denser parts of the loop (due to natural gravitational stratification) the aforementioned high percentage is no longer needed. It has been known that, in homogeneous plasma, when perpendicular wavelength of Alfven wave (AW) approaches kinetic scales such as e.g. ion-inertial length, it acquires magnetic-field-aligned (parallel) electric field, which can efficiently accelerate electrons \citep{2000SSRv...92..423S}. Further, \citet{2005A&A...435.1105T}
have shown that if DAW propagates in plasma with transverse (with respect to external magnetic field) density inhomogeneity, the generated parallel electric field is orders of magnitude higher than (i) homogeneous plasma case and (ii) Dreicer electric field (one that triggers electron run-away acceleration). Subsequently 
\citet{dt12}
 has revisited the problem with full 3D particle-in-cell approach. \citet{ofman10} considered
similar set up as in \citet{2005A&A...435.1105T} but instead of considering one DAW harmonic with $0.3 \omega_{\rm { cp}}$ he considered $f^{-1}$ AW spectrum and added ${\rm He^{++}}$ ions and used Hybrid simulation model. 
Here $\omega_{\rm { cp}} = e B / m_{\rm p} $ is { proton} cyclotron frequency.
Note that our approach uses PIC code so it can resolve electron-scale physics contrary to Ofman who used a Hybrid code, which can resolve only ion-scale physics. Now in the present work we essentially revisit Ofman's set up run it for two cases:
1. when transverse density gradient is $\approx 4 c/\omega_{\rm { pe}}$ (as in 
\citet{2005A&A...435.1105T,dt12}), i.e. on "electron"-scale;
2. when the gradient is on { proton} scale circa $\approx 40 c/\omega_{\rm { pe}}$; In this paper
 novel numerical simulation results are presented. Including the scaling of the magnetic fluctuations power spectrum steepening in the higher density regions, and the heating channelled to these regions from the surrounding lower 
density plasma due to wave refraction. We also present runs where DAW collide multiple times in a situation similar to  \citet{doi:10.1063/1.4812805,daif22}. 
{ We stress that despite the similarities between works of 
\citet{ofman10} and \citet{2005A&A...435.1105T,dt12}, including the 
current work, the former used physical parameters for solar wind, while
the latter use physical parameters for solar coronal flaring loop tops.
Therefore direct comparison between results and magnetic fluctuation 
spectra scaling laws is not appropriate.
In this context, also we note that subsequent studies exist: (i) \citep{ofman15} which
 performed 2.5D hybrid simulations to investigate the
role of initially imposed broad-band wave spectra in a drifting
and expanding solar wind plasma and 
(ii) \citep{ofman17} which study the effects of inhomogeneous (across the magnetic field) background streaming focusing on the fast solar wind. 
They explore the effects of an initial relative, inhomogeneous ion drift on the perpendicular ion heating and cooling and consider the effects of solar wind expansion. }

It should be noted the above mentioned electron fraction/number 
problem is not without
a debate. \citet{Kontar_2023} argue that 
for a solar flare on September 10, 2017, their observations show 
100 times smaller fraction of accelerated electrons, i.e. just 
mere $\approx 10^{-2}$,
compared to {\it the same flare} observation in a different 
observing wavelength by 
 \citep{2020Sci...367..278F,2022Natur.606..674F},
who on contrary, claim that nearly 100\% electrons need to be accelerated
to match the observations.
\citet{Kontar_2023}'s explanation for the small fraction
is a steady resupply of electrons to the acceleration site.
Since the present paper is purely theoretical and is based
on numerical simulations, we do not make a judgement which
of the aforementioned observations is correct, while insisting that
electron acceleration by DAWs is  a viable mechanism to
accelerate electrons from loop top to foot-points. Nor we aim to
match or model particular observational data set, staying
within bounds of a theoretical, "the first principles" approach. 

To summarize the background work and emphasize
the novelty of this work, compared to our previous results, \citep{2005A&A...435.1105T,dt11,dt12}) is 4-fold:
(i) instead of exciting DAWs at the left edge of the domain which mimics
loop top we drive DAWs in the middle of the domain, which
more realistically represents DAWs generated at the loop top;
(ii) we consider wide spectrum of DAWs as specified in Table \ref{t1};
(iii) we added ion-cyclotron resonant ${\rm He^{++}}$ species;
(iv) we considered cases when DAW collide multiple times.

\section{The model}

We use 2.5D version of EPOCH which is a
multi-dimensional, fully electromagnetic, PIC code
\citep{Arber:2015hc}.
In EPOCH code physical quantities are in
SI units. In this paper we use the 
normalization for the graphical
presentation of the results as follows:
Time and distance are shown in $\omega_{\rm pe}^{-1}$
and $c / \omega_{\rm pe}$. 
Electric and magnetic fields are shown in units of
 $E_{x0}=\omega_{\rm pe}c m_e /e$ and  
 $B_{x0}=\omega_{\rm pe} m_e /e$, respectively.
Here $\omega_{\rm pe}=[n_{e0} e^2/(m_e \varepsilon_0)]^{0.5}$
is electron plasma frequency.
$n_{e0} =10^{16}$ m$^{-3}$ is electron number density in the
lowest density 
regions of the simulation 
domain ($y=0$ and $y=y_{\rm max}$).
This sets electron plasma frequency at 
 $\omega_{\rm pe} = 5.64 \times 10^9$ Hz radian.
The size of simulation box is either
as $x=5000$ and $y=200$ grid points  
when transverse density gradient is 
$\approx 4 c/\omega_{\rm { pe}}$
(electron scale);
or $x=5000$ and $y=2000$ grid points 
when transverse density gradient is $40 c/\omega_{\rm { pe}}$ (ions scale).
{ The grid size in all numerical simulations presented in this
work is electron Debye radius 
$\lambda_D = v_{\rm th,e}/ \omega_{\rm pe}$.
Here $v_{\rm th,e}=\sqrt{k_B T/m_e}$ is an electron thermal velocity.}
The plasma beta in this study is fixed at 
$\beta= 2 (v_{th,i}/c)^2(\omega_{\rm { pp}}/\omega_{\rm { cp}})^2 = 
n_0(0,0)k_B T /(B_0^2/(2\mu_0))=0.02$. Thus $\beta=0.006746 < m_e/m_i=1/16=0.0625$, which means that we are so-called Inertial Alfven wave
(IAW) regime. 
According to \citet{2000SSRv...92..423S}
nomenclature is such that 
in the different regimes inertial ($\beta < m_e/m_i$) and kinetic 
($\beta > m_e/m_i$). The latter waves are called Kinetic Alfven waves (KAW).
We could computationally afford somewhat unrealistic
electron to proton mass ratio of  $m_e/m_i=1/16$, because our largest run
on $x=5000$ and $y=2000$ grid points with end-simulation-time of
$t_{\rm end} = 375.0/\omega_{\rm { cp}}$ takes 
1 day, 1 hour, 18 minutes on 2048 CPU cores (16 x 128 core nodes) using 
Dell PowerEdge R6525 compute nodes each with 2 x AMD EPYC 7742 (Rome) 2.25 GHz 64-core processors i.e. 128 cores per node.
{ We admit that indeed the mass ratio $m_e/m_i=1/16$
is an unrealistic one compared to the actual $m_e/m_i=1/1836$ for the case
of a proton. 
The mass ratio of
 $m_e/m_i=1/16$ sets the ratio of Alfven speed to speed of light of 
$V_A/c=\omega_{\rm cp}/\omega_{\rm pp}=0.25$, which is
{ larger} than the realistic one of $V_A/c=0.023$.
However, many studies use this simplification to speed up
the numerical computation. Our previous study \citet{Tsiklauri_2007} has considered the effect of variation of the mass ratio and
found 
that amplitude attained by the generated 
$E_\parallel$ decreases linearly as the
inverse of the mass ratio $m_i/m_e$, i.e. $E \propto 1/m_i$. This result means that
 for a realistic mass ratio of $m_i/m_e = 1836$ our
empirical scaling law produces $E_\parallel = 14$ V$/$m for solar coronal parameters, which is a significant value for electron acceleration.}
In the present simulation
we use 100 electrons, 100 protons and 100 He$^{++}$ ions
per cell which means that for 
there are total of $3\times10^8$ particles in the simulation i.e.
$1\times10^8$ per species for the $5000 \times 200$ spatial grid
case. When we use $5000 \times 2000$ grid there are obviously 10 more 
particles i.e. total of $3\times10^9$ particles and $1\times10^9$ 
particles per species.

We impose constant background magnetic field of
strength $B_{0x} =0.032075$ Tesla along
$x$-axis. This corresponds to $B_{0x}=1.0 (\omega_{\rm pe} m_e /e)$.
This means that 
with the normalization used for the visualization
purposes, normalized background magnetic field is unity. 
Also such choice sets the ratio of electron
cyclotron and plasma frequencies as unity $\omega_{\rm ce}/\omega_{\rm pe} = 1.00$.
Electron, proton (ion) and  He$^{++}$ temperature at the simulation box edge
i.e. in the homogeneous density parts of the domain 
are fixed at $T_e(0,0)=T_i(0,0)= T_{\rm He^{++}}(0,0)\equiv T_0=2\times10^7$K. 
This temperature along  with $n_{e}(0,0) \equiv n_{e0}
 =10^{16}$ m$^{-3}$,
sets plasma parameters of a typical dense solar coronal 
flaring loops. 
In order to keep charge neutrality the relation
between charges species is $1.0 n_{e0}=0.9 n_{i0} +0.05 n_{\rm He^{++}0}$, because He is twice ionized. Here subscript $0$ refers to location $(0,0)$
i.e. edge of the domain here density is uniform. 

We note that the results presented here are equally applicable to
both magnetospheric auroral density cavities 
\citep{genot99,genot00,tom_lys96,daif17,daif22}
and
solar coronal loops
\citep{2005A&A...435.1105T,Tsiklauri_2007,dt11,dt12,dt16}, 
as far as there is a density gradient
present, be it negative (density depletion)
in the case of the cavities or
positive (i.e. density enhancement)
in the case of solar coronal loops.
{  Discussing magnetosphere analogy here sets an appropriate
interdisciplinary context of the current study. We note that only in the Earth magnetosphere DAWs are currently observed in situ,
as the Sun is too far away from DAWs to be detected remotely.}

We model density variation 
of all plasma constituent particle 
species, across
the background magnetic field, of a 
solar coronal loop as
\begin{multline}
{n_{\rm He^{++}}(y)}/n_{\rm He^{++}0}=
{n_i(y)}/n_{i0}=  
{n_e(y)}/n_{e0}=\\
=1+3 \exp\left[-\left(\frac{y-
y_{\rm max}/2.0}{0.25 y_{\rm max}}\right)^6\right]
\equiv f(y).
\label{eq1}
\end{multline}
Eq.(\ref{eq1}) shows that in the middle part
along y-coordinate
the number density is 
increased by a factor of four
compared to edges on either side.
Also this density profile has two
density gradients 
in the vicinity of 
$y=0.25 y_{\rm max}$ and $y=0.75 y_{\rm max}$, both
having a width of about $0.2 y_{\rm max}={ 4 c / \omega_{pe}}$.
The background temperature of positive ions and electrons 
is varied as
\begin{equation}
{T_{\rm He^{++}}(y)}/{T_0}=
{T_{i}(y)}/{T_0}=
{T_e(y)}/{T_0}=f(y)^{-1}.
\label{eq2}
\end{equation}
This makes sure that the 
thermal pressure in all plasma particle species 
is constant across y-coordinate. Therefore, because
$B_{x0}=const$,
such initial conditions have
the total pressure (the sum of thermal and magnetic pressures) being 
constant.
{ A graphical visualization of the density and
temperature profiles according to
Equations \ref{eq1} and \ref{eq2}  have been published before by the
author and as a further reference can be found as
figure 2 from \citet{dt16}.}

When we consider a single harmonic, 
we trigger generation of dispersive Alfven waves by
driving the middle part in x-coordinate i.e.
$x=x_{\rm max}/2$, for all y-s, in the following manner 
\begin{multline}
E_y(x=x_{\rm max}/2,y,t+\Delta t)= E_y(x=x_{\rm max}/2,y,t)-     \\
A_y\sin(\omega_d t)\left(1-\exp\left[-(t/t_0)^2\right]\right), 
\label{eq3}
\end{multline}
\begin{multline}
E_z(x=x_{\rm max}/2,y,t+\Delta t)=E_z(x=x_{\rm max}/2,y,t)-     \\
A_z\cos(\omega_d t)\left(1-\exp\left[-(t/t_0)^2\right]\right).
\label{eq4}
\end{multline}
Such driving generates  
{\it left-hand}-polarized DAWs which then travel
from the middle part of the domain to the left ($x=0$) and right ($x=x_{\rm max}$) edges. 
We chose left-hand-polarization for ion cyclotron resonance to occur for 
${\rm He^{++}}$ species.
Note that such DAW driving is more realistic
to solar flare conditions with flare taking place
at loop top, i.e. at $x=x_{\rm max}/2$ than compared to earlier
cases \citep{2005A&A...435.1105T,Tsiklauri_2007,dt11,dt12},
where the driving was taking place at the 
domain's left edge, $x=0$, and the DAWs where appearing
on the right edge, $x=x_{\rm max}$, due to periodic 
boundary conditions. 
In our present numerical runs the boundary conditions
are periodic for every physical quantity be it
particles of fields.
In equations (\ref{eq3}) and (\ref{eq4}), 
$\omega_d$ is the DAW driving frequency,
 fixed at $\omega_d=0.3\omega_{\rm { cp}}$.
With such 
driving frequency, there is no significant proton-cyclotron damping  and also the generated DAW is Alfvenic in the frequency sense.
The onset time of the driver, $t_0$ is 
fixed at $3.0 \omega_{\rm { cp}}^{-1}$
i.e.  $48.000  \omega_{\rm pe}^{-1}$ for the case of $m_i/m_e=16$. 
This means that the driver onset time is about $3^2\omega_{\rm { cp}}^{-1}
(=9 \omega_{\rm { cp}}^{-1})$. 
{ This means that 
after the driver onset time 
we constantly inject DAWs with amplitude of $0.05 B_0$, i.e. 
5\% of the background magnetic field.}
The initial amplitudes of the $E_\perp$ are chosen 
to yield  circular polarization
$A_y=A_z=0.05(cB_{x0})$. This corresponds to  
the relative DAW amplitude 5\% of the background magnetic field 
strength.
{ This value of the amplitude is similar to that considered by \citet{ofman10}
who select 3\% of the background magnetic field.
We would like to comment that both studies suffer from the same shortcoming of
{\it too large} value of $\delta B/B_0$ of $3-5$\%, which
is otherwise appropriate for many MHD scale waves, as observed extensively in the corona 
(e.g. \citet{2014SoPh..289.3233L}). The above MHD waves are 8-10 orders of magnitude longer than the 
DAW discussed in the present paper or the ones
discussed in \citet{ofman10}.
However, this magnetic perturbation ratio and corresponding magnetic energy 
flux is unrealistically large for kinetic scale DAWs be it on ion or electron scale.  
Evidence of the observed magnetic fluctuation spectra throughout 
the heliosphere shows strong power law decrease of the energy with 
frequency (e.g., \citet{2021PhRvE.103f3202A}, \citet{Lotz_2023}). 
The steep scaling of the power spectrum with $k$ was also demonstrated 
in the present model (Figure \ref{fig3}) (see below). Therefore, the 
assumption of $\delta B/B_0$ of 3-5\% for these high 
frequency kinetic scale waves or DAWs is {\it unrealistic}.
The {\it main reason} both PIC (this study) and Hybrid (\citet{ofman10})
consider few percent amplitude waves is because considering 8-10 order of magnitude
smaller amplitude is not possible because any kinetic model suffers from so-called
'shot noise'. Kinetic models typically consider 100 or 100s of particles per cell which is 
many orders of magnitude smaller than the realistic number of particles per volume.
The level of 'shot noise' scales as $\propto 1 / \sqrt{N}$, where $N$ is number of
particles per cell. In order to reduce amplitude we need to reduce 'shot noise',
but that is computationally 
impossible because it would require many billions of particles per cell.}
{ However we remark that 
in the solar wind the magnetic field is 6 orders of magnitudes weaker that the magnetic field in typical active region loops, and the relative amplitudes of DAWs is expected to be orders of magnitude larger in the solar wind compared to flaring loops.}

When we consider a broadband $1/f$ spectrum, as in \citet{ofman10},
the following DAW driving is imposed at 
the middle part in x-coordinate,
$x=x_{\rm max}/2$:  
\begin{multline}
E_y(x=x_{\rm max}/2,y,t+\Delta t)= E_y(x=x_{\rm max}/2,y,t)-     \\
\sum_{i=1}^{N}i^{-0.5} F A_y 
\sin \left(\left[\omega_{\rm min}+
\frac{(\omega_{\rm max}-\omega_{\rm min})(i-1)}{N-1}\right] t
+0.05 \pi R(i) \right)\\
\times \left(1-\exp\left[-(t/t_0)^2\right]\right), 
\label{eq5}
\end{multline}
\begin{multline}
E_z(x=x_{\rm max}/2,y,t+\Delta t)=E_z(x=x_{\rm max}/2,y,t)-     \\
\sum_{i=1}^{N}i^{-0.5} F A_z
\cos\left(\left[\omega_{\rm min}+
\frac{(\omega_{\rm max}-\omega_{\rm min})(i-1)}{N-1}\right] t
+0.05 \pi R(i)\right)\\
\times \left(1-\exp\left[-(t/t_0)^2\right]\right).
\label{eq6}
\end{multline}
Here $N$ is the number modes which is varied as $N=128,512$.
$R(i)$ are 128 or 512 random numbers between 0 and 1 which 
stay the same throughout the simulation and they mimic
some randomness introduced in the sine
 and cosine phase of the order of 
5\%, as in \citet{ofman10}.
See Table~\ref{t1} for details.
DAW driving frequency is in the range of 
$0.3-0.6\omega_{\rm { cp}}$, i.e. in Eqs.\ref{eq5} and \ref{eq6}
$\omega_{\rm min}=0.3\omega_{\rm { cp}}$ and $\omega_{\rm max}=0.6\omega_{\rm { cp}}$.
This means that for protons our numerical simulation
runs always stay clear from { proton} cyclotron resonance because in
our notation 
$\omega_{\rm { cp}}=
 e B /m_{\rm p}$ denotes {\it proton} cyclotron frequency.
Note that for ${\rm He^{++}}$ ions, which has 2 positive charges and has a mass
of 4 nucleons (2 protons and 2 neutrons),
cyclotron frequency is $\omega_{\rm He^{++}}=2 e B /(4 m_{i})=
0.5 e B /m_{\rm p}=0.5 \omega_{\rm { cp}}$, which is near the middle ($0.45\omega_{\rm { cp}}$) of
DAW frequency range of $0.3-0.6\omega_{\rm { cp}}$.
Therefore we intentionally have the effect of ${\rm He^{++}}$
ion cyclotron resonance heating as in \citet{ofman10},
which, in turn, was motivated by observations of 
high perpendicular temperature observations of heavy ion
species such as e.g. ${\rm He^{++}}$ \citep{PhysRevLett.110.091102} and
O VI \citep{Telloni_2007}. 
In the near future, there will be significant discoveries in this field
with the advent of the Solar Orbiter Heavy Ion Sensor (HIS), which is a time-of-flight ion mass spectrometer dedicated to measuring heavy ions in the solar wind with unprecedented precision \citep{HIS2023}.

\begin{table}
\caption{Table of numerical runs considered.}
\centering
\begin{tabular}{lccr} 
\hline
Run & $t_{\rm end} \omega_{\rm { cp}}$ & Harmon. & Brief Description \\
\hline
1 & 75 & 1 &   Eqs.\ref{eq3}-\ref{eq4}\\
2 & 75 & 512 &  Eqs.\ref{eq3}-\ref{eq4}, the sum with $A_{y,z} \to A_{y,z}/ 512$  \\
3 & 75 & 128 &  Eqs.\ref{eq5}-\ref{eq6}, with $F$ from Eq.\ref{eq7}\\
4 & 75 & 512 &  Eqs.\ref{eq5}-\ref{eq6}, The Main Run\\
5 & 750 & 512 & Eqs.\ref{eq5}-\ref{eq6}, 10x larger $t_{\rm end}$\\
6 & 375 & 512 &  Eqs.\ref{eq5}-\ref{eq6}, 10x larger $d n/ dy\simeq 40 c/\omega_{\rm { pe}}$\\
7 & 75 & 512 &  homogeneous version of run 4 \\
8 & 750 & 512 & homogeneous version of run 5\\
9 & 375 & 512 & homogeneous version of run 6\\
\hline
\end{tabular}
\label{t1}
\end{table}
The factor $F$ in Eqs.\ref{eq5}-\ref{eq6} is calculated using 
\begin{multline}
F = \frac{1}{N }  
\sum_{i=0}^{N-1} \left[\frac{ \omega_{\rm min}+
(\omega_{\rm max}-\omega_{\rm min})\;i}{(N-1)\omega_{\rm min}}\right]^2
\Biggl/\left(\sum_{i=0}^{N-1}(i+1)^{-0.5}\right).
\label{eq7}
\end{multline}
For $N=128$ $F=0.11006650$ and $N=512$ $F=0.053259745$.
Such choice insures that the broadband spectrum of
DAWs delivers the same power as one alters number of harmonics, which will be demonstrated below. 

Run 1 from Table \ref{t1} uses just one harmonic as per Eqs.\ref{eq3}-\ref{eq4}
with the end simulation time of $t_{\rm end} \omega_{\rm { cp}}=75$. This is
essentially reproduction of a previous run considered in \citet{dt11}
for the reference purposes and its input is used in production of Fig.\ref{fig4}.

Run 2 from Table \ref{t1} uses  Eqs.\ref{eq3}-\ref{eq4}
with {\it the sum} of 512 harmonics,
but each with amplitude 512 times smaller than Run 1. Here 
the end simulation time is $t_{\rm end} \omega_{\rm { cp}}=75$. This is
essentially reproduction of a previous run considered in \citet{dt11}
for reference purposes and its input is used in production of Fig.\ref{fig4}.
The purpose of this run is to ascertain that using the sum of 512 single
frequency 
harmonics with amplitudes 512 times smaller than Run 1 produces identical
results.

Run 3 from Table \ref{t1} uses 128 harmonics as per Eqs.\ref{eq5}-\ref{eq6}
with $F$ from Eq.\ref{eq7}
with the end simulation time of $t_{\rm end} \omega_{\rm { cp}}=75$. 
The purpose of this this run is to ascertain changing number of harmonic
still delivers the same power compared to 
Run 4 (512 harmonics).

Run 4 from Table \ref{t1} uses 512 harmonics as per Eqs.\ref{eq5}-\ref{eq6}
with $F$ from Eq.\ref{eq7}
with the end simulation time of $t_{\rm end} \omega_{\rm { cp}}=75$.
This the regarded as "the main run" in a sense that: 
(i) it shows the differences brought by wide spectrum of DAWs;
(ii) most detailed simulation results will be shown for this run.

Run 5 from Table \ref{t1} has the end simulation time of $t_{\rm end} \omega_{\rm { cp}}=750$. Otherwise it is identical to the main run (Run 4).
The purpose of this run is to consider situation when DAWs collide
multiple times. In numerical runs with $t_{\rm end} \omega_{\rm { cp}}=75$
DAWs travel from the domain middle part in x-coordinate i.e.
$x=x_{\rm max}/2$ towards $x=0$ and $x=x_{\rm max}$, {\it nearly reaching}
the said boundaries. Because of periodic boundary
conditions DAWs will re-enter from the domain opposite sides, as the time
progresses. 
This means that for Run 5
$t_{\rm end} \omega_{\rm { cp}}=750$ with 10-times longer simulation time
the DAWs will collide at least 
9 times, near stopping short of the 10th collision
at $t_{\rm end} \omega_{\rm { cp}}=750$.
With Run 5 we wish to study how multiple DAW frontal collisions
will affect particle acceleration in a situation similar to
\citet{doi:10.1063/1.4812805,daif22}.

Run 6 from Table \ref{t1} has 
the length-scale of the density gradient in $y$-direction increased
10 times to nearly "ion"-scales
$d n/ dy\simeq 40 c/\omega_{\rm { pe}}$. 
This is achieved by setting 10-times the number of grids in $y$-direction, i.e.
$x=5000$ and $y=2000$ grid points in each direction.
Otherwise it is identical to the main run (Run 4).
The purpose of this run is to study how particle acceleration is
affected by up-scaling the transverse density gradient from
electron scales $d n/ dy\simeq 0.2 y_{\rm max}= 
4 c/\omega_{\rm pe}$ to ion-scales $d n/ dy\simeq 0.2 y_{\rm max}= 
40 c/\omega_{\rm { pe}}$.
In Run 6 we set $t_{\rm end} \omega_{\rm { cp}}=375$, the
maximal possible with the computational resources available.
This run took 1 day, 34 minutes on 2048 CPU cores (16 x 128 core nodes) using 
Dell PowerEdge R6525 compute nodes.

Runs 7,8,9  from Table \ref{t1} are homogeneous versions of Runs 4,5,6 respectively.
Note the different end simulation times used in each run.
The purpose of these numerical runs is to quantify
the difference in properties of particle acceleration in the
homogeneous and inhomogeneous cases respectively.

\section{The Results}

\begin{figure*}
\begin{center}
  \makebox[\textwidth]{\includegraphics[width=1.745\columnwidth]{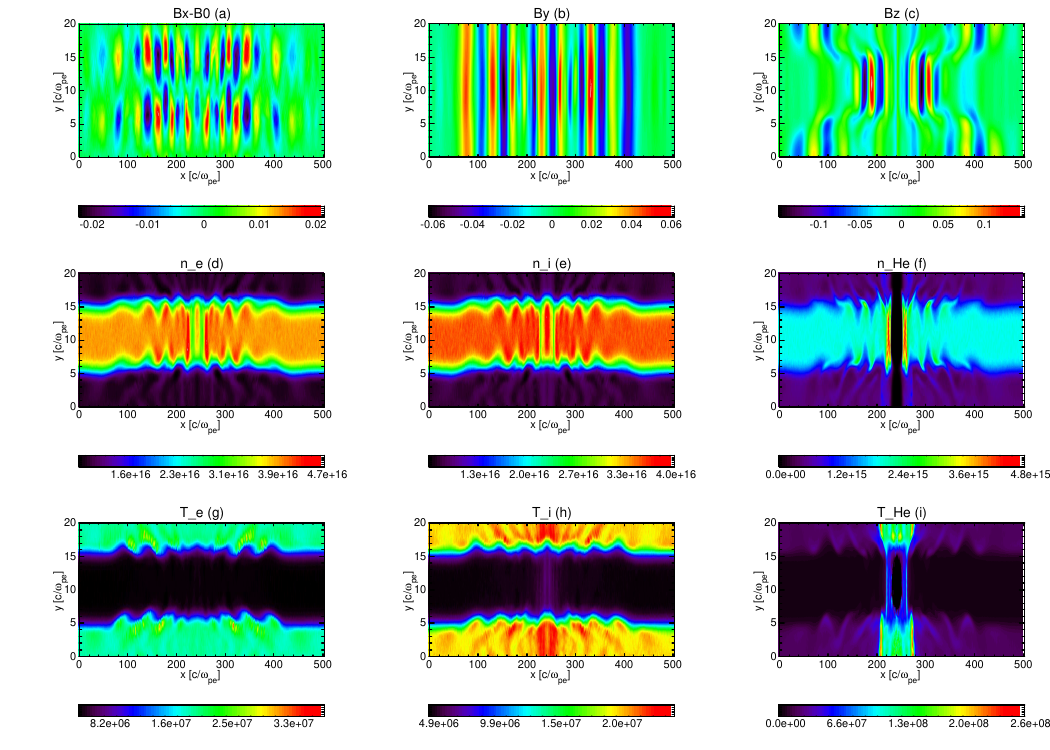}}
\end{center}
\caption{A snapshot of $B_x-B_{x0}$, $B_y$, $B_z$, 
$n_e$, $n_i$, $n_{\rm He^{++}}$, $T_e$, $T_e$, $T_{\rm He^{++}}$,
at $t=75 / \omega_{\rm { cp}}$ for Run 4. See Table \ref{t1} for details.}
    \label{fig1}
\end{figure*}

In Fig.\ref{fig1} we show a snapshot of $B_x-B_{x0}$, $B_y$, $B_z$, 
$n_e$, $n_i$, $n_{\rm He^{++}}$, $T_e$, $T_e$, $T_{\rm He^{++}}$,
at $t=75 / \omega_{\rm { cp}}$.
We gather from this plot that the perturbation
of magnetic field along the background
magnetic field is generated as shown in panel (a). It closely
follows pattern of the generated background magnetic filed-aligned
electric field $E_x$ (not shown for brevity), i.e.
the generated field-aligned $E_x$ and $B_x-B_{x0}$ have
wave-like structure confined to the density gradient across
the magnetic field with amplitudes $\approx 2\%$. 
Similarly as in previous results
$B_y$ (panel (b))
is not phase-mixed while $B_z$ (panel (c)) shows a clear pattern of phase mixing
when the Alfvenic wave (DAW to be precise) travels slower in the over-dense
region $5  c/\omega_{\rm pe} < y< 15 c/\omega_{\rm pe}$ as
Alfven speed is proportional to the inverse square root of the
plasma density.
In fact, panels (a),(b),(c) look very similar to panels (a),(b),(c) in
Fig.1 from \citet{dt11}, in that amplitudes are similar, and phase-mixing
pattern looks the same. However, the notable difference is that in
Fig.\ref{fig1} now (i) instead of exciting DAWs at the left edge of the domain  now we use more realistic DAWs generated at the solar coronal loop top, i.e.
$x=x_{\rm max}/2$ and 
(ii) we consider wide spectrum of DAWs as specified in Table \ref{t1}.
From panels (d) and (e) we gather that electron and { proton}
 number densities
show significant transverse to the background magnetic field
perturbations. 
We see in panels (g) and (h) the electron and { proton} temperatures
also show wake like perturbation confined near density gradient
regions and these gradient regions are the locations of a noticeable
temperature increase. 
We note again that panels (d),(e),(g),(h) look very similar to panels (a),(b),(c),(d) in
Fig.2 from \citet{dt11}, with the proviso of the middle of the domain
driving and wide spectrum.
\citet{dt11} did not consider ${\rm He^{++}}$ hence the panels (f) and (i)
which show ${\rm He^{++}}$ number density and temperature respectively, are new.
${\rm He^{++}}$ number density shows (panel (f)) significant non-linear,
cat-claw like perturbations with the maximum concentrated
near the location DAW driving at $x=x_{\rm max}/2$.
${\rm He^{++}}$ temperature (panel (i) shows
 significant
increase near the location DAW driving.
This is due to the fact that for ${\rm He^{++}}$ ion cyclotron condition is
met.

\begin{figure*}
\begin{center}
  \makebox[\textwidth]{\includegraphics[width=1.745\columnwidth]{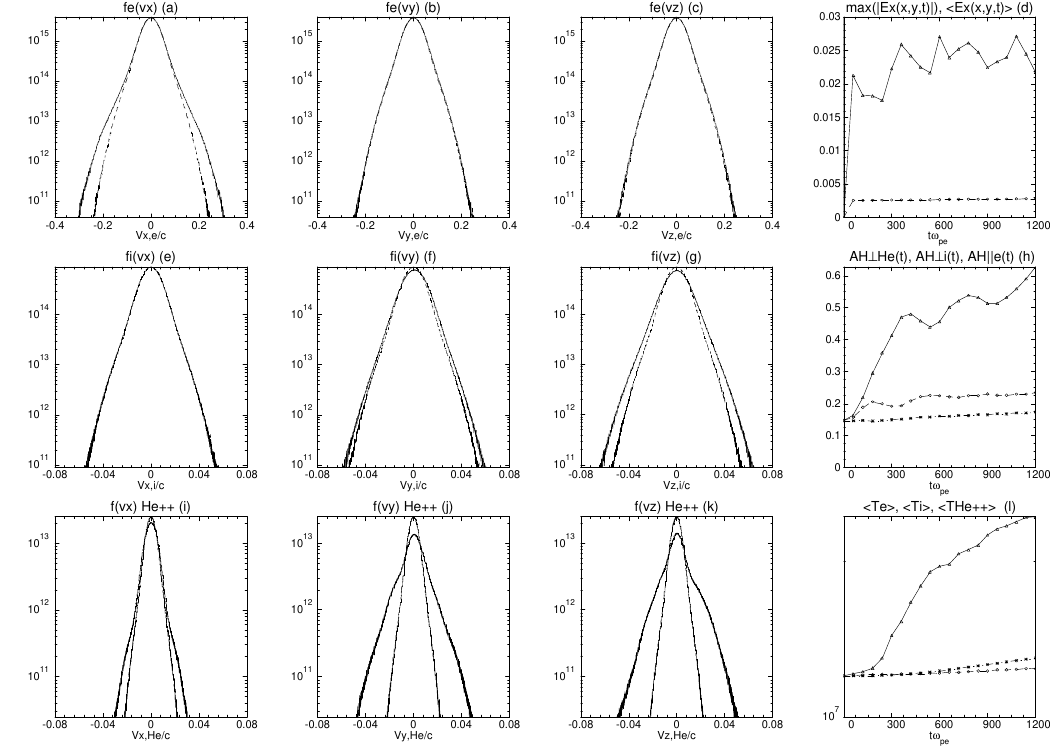}}
\end{center}
    \caption{Distribution function dynamics for
electron $V_x$, $V_{ y}$,$V_z$ components (panels (a), (b), (c));
ion (proton) $V_x$, $V_{ y}$,$V_z$ components (panels (e), (f), (g));
${\rm He^{++}}$ $V_x$, $V_{ y}$,$V_z$ components (panels (i), (j), (k));
In panels (a), (b), (c), (e), (f), (g), (i), (j), (k),
the inner dashed curve show the distribution
function (distribution by velocities) at $t=0$, while
outer solid curve at the final simulation time.
In panel (d) we show ${\rm max(|E_x/E_{x0}|)}$ (solid line with triangles)
and ${\rm <|E_x(x,y,t)/E_{x0}|>}$ (dashed line with diamonds).
In panel (h) 
solid line with triangles is for $AH_{\perp, He}(t)$;
dashed line with diamonds is for $AH_{\perp,i}(t)$;
dash-dotted line with stars is for $AH_{\parallel,e}(t)$.
In panel (l) we show ${\rm He^{++}}$, proton and electron temperature
evolution in time.
Solid line with triangles is for ${\rm <T_{He^{++}}>}$;
dashed line with diamonds is for ${\rm <T_{i}>}$;
dash-dotted line with stars is for ${\rm <T_{e}>}$.
See averaging definitions in the text.
All panels show data for the Run 4.}
    \label{fig2}
\end{figure*}

Fig.\ref{fig2} has a purpose to quantify how particle acceleration
main features, fractions and the budget, 
split between different plasma species,
is modified by the introduction of 
$f ^{-1}$ frequency spectrum of DAWs with driving frequency in the range $0.3-0.6\omega_{\rm { cp}}$. Thus, in Fig.\ref{fig2} we show
distribution function dynamics for
electron $V_x$, $V_{ y}$,$V_z$ components (panels (a), (b), (c));
ion (proton) $V_x$, $V_{ y}$,$V_z$ components (panels (e), (f), (g));
${\rm He^{++}}$ $V_x$, $V_{ y}$,$V_z$ components (panels (i), (j), (k));
In panels (a), (b), (c), (e), (f), (g), (i), (j), (k),
the inner dashed curve show the distribution
function (distribution by velocities) at $t=0$, while
outer solid curve at the final simulation time.
Note that for protons and ${\rm He^{++}}$ the
inner dashed curves for $t=0$ look almost as solid due to
large number of data points (each distribution function is
resolved with 100,000 points in all panels).
By comparing panels (a), (b), (c), (e), (f), (g)
with similar run e.g. panels 
(a),(b),(c),(d),(e),(f)
in Fig.3 from \cite{dt11}
we gather many similarities.
In particular we see in panel (a) that electron beams 
are formed
near the DAW phase velocity. A study of
\citet{dt12} has shown that the electron beam formation
along the uniform background magnetic field
is due to Landau resonance between electrons
and DAW. In fact \citet{dt12} showed if one varies
ion(proton) mass, which prescribes DAW phase speed,
 then location of the beam peak in the distribution function
changes accordingly, 
signalling occurrence of the Landau resonance.
In panels (b) and (c) we see that distribution functions
with respect to $V_y$ and $V_z$ at time zero and the final simulation 
time do overlap. Hence, we deduce from these panels that no sizable
electron heating in the perpendicular to the
magnetic field direction is seen.
While, for { protons} we see
in corresponding panels (f) and (g) that primarily perpendicular to the
magnetic field broadening of the distribution occurs
due to DAW energy injection/driving.
No change in the { proton} parallel to the magnetic
field distribution function (panel (e)) is seen, suggesting
that on this time-scale { proton} parallel dynamics is not affected
by DAW driving.
In panel (d) we show ${\rm max(|E_x/E_{x0}|)}$ (solid line with triangles)
and ${\rm <|E_x(x,y,t)/E_{x0}|>}=\sum_{i=1,j=1}^{i=n_x,j=n_y} |E_x(i,j,t)/E_{x0}|/(n_x \times n_y)$ (dashed line with diamonds),
i.e. maximal and average of the absolute values
of electric field x-component.
In all line plots when line is over-plotted with a symbol (e.g. such as
triangles or diamonds), the symbol fixes a 
moment in time when such value
was recorded.
We note that panel (d) is similar to panel (g) in Fig.3 from \cite{dt11}
in that ${\rm max(|E_x/E_{x0}|)}$ attains similar values of 
$\approx 0.03$. As noted before by \citet{dt11} for 
solar flaring plasma parameters this 
exceeds Dreicer electric field, which is
a condition for run-away electron acceleration,
 by eight orders of magnitude.
In order to quantify acceleration of particles we as in previous
works \citet{dt11,dt12}, make use of the following quantities/indexes:
\begin{equation}
AH_{\parallel,e}(t)=
\frac{\int_{|v_{x}| > \langle v_{th,e}(t)\rangle}^\infty      
f_e(v_x,t)dv_x\Biggl/\left(2 L_{IH,y}\times L_{x,max}\right)}
{ \int_{-\infty}^\infty   f_e(v_x,0) dv_x\Biggl/\left( L_{y,max}\times L_{x,max}\right)},
\label{eq8}
\end{equation}
\begin{multline}
AH_{\perp,i, He}(t)= \\
\frac{\int_{|v_{\perp}| > \langle v_{th,i, He}(t)\rangle}^\infty      f_{i,He}(v_\perp,t)dv_\perp\Biggl/\left(2 L_{IH,y}\times L_{x,max}\right)}
 { \int_{-\infty}^\infty   f_{i,He}(v_\perp,0) dv_\perp\Biggl/\left( L_{y,max}\times L_{x,max}\right) },
\label{eq9}
\end{multline}
where $f_{e,i,He}$ are electron,  proton
and H++ velocity distribution functions and
$<>$ brackets denote average over y-coordinate, because temperature and density vary across y-coordinate.
$L_{IH,y}$ is the width of each density gradients according to Eq.\ref{eq1}.
$L_{y,max}$ is the full width in the $y$-direction. The 
numerator of Eq.(\ref{eq8}) yields 
number of super-thermal electrons in the parallel to the
magnetic field direction, 
divided by the area ($2 L_{IH,y}\times L_{x,max}$) in the region where 
particle acceleration takes place. 
Note that in this study because we consider some runs 
that are 10 longer in time, than previously,
and hence expect commensurately a larger amount of
energy being injected into the physical system due to
continuous input from DAW wave driving, temperature of
each plasma species is expected to increase.
Therefore, {\it crucial difference}  from the
previous definitions as in \citet{dt11,dt12},
 $v_{th,e}(t)$ and $v_{th,i, He}(t)$
are now {\it functions of time}.
Equations \ref{eq8} and \ref{eq9} provide 
fraction (i.e. percentage) 
of super-thermal plasma species parallel and
perpendicular to the regular magnetic field.
Thus, in panel (h) (of Figure \ref{fig2})
solid line with triangles is for $AH_{\perp, He}(t)$;
dashed line with diamonds is for $AH_{\perp,i}(t)$;
dash-dotted line with stars is for $AH_{\parallel,e}(t)$.
We gather from panel (h) that, 
within $t_{\rm end} \omega_{\rm { cp}}=75$,
 $AH_{\perp, He}(t)$ attains value of 0.63, i.e.
63\% of ${\rm He^{++}}$ become super-thermal.
As seen in panels (j) and (k)
this is mainly due to the ion-cyclotron resonance in the
perpendicular to $B_{x0}$ direction. So it is ${\rm He^{++}}$ 
heating, not acceleration as such.
A similar behavior for $AH_{\perp,i}(t)$ 
attaining values of 23\%
is seen for protons in panel (h)
i.e. by looking at panels (f) and (g) we
see perpendicular proton heating, but not as dramatic 
as for ${\rm He^{++}}$ because cyclotron resonance condition 
is not met for protons. 
In panel (h) we note that $AH_{\parallel,e}(t)$
attains even smaller value so fraction of super-thermal
electrons is 17\%.
Compared to panel (h) from Fig.3 in \citet{dt11},
now $AH_{\perp,i}(t)$  and $AH_{\parallel,e}(t)$
attain smaller values because we inject the same 
amount of power but most of it goes to ${\rm He^{++}}$ due to
ion-cyclotron resonance.
In panel (l) we show ${\rm He^{++}}$, proton and electron temperature
evolution in time.
In particular,
solid line with triangles is for ${\rm <T_{He^{++}}>}=\sum_{i=1,j=1}^{i=n_x,j=n_y} 
T_{He^{++}}(i,j,t)/(n_x \times n_y)$;
dashed line with diamonds is for ${\rm <T_{i}>}=\sum_{i=1,j=1}^{i=n_x,j=n_y} 
T_{i}(i,j,t)/(n_x \times n_y)$;
dash-dotted line with stars is for ${\rm <T_{e}>}=\sum_{i=1,j=1}^{i=n_x,j=n_y} 
T_{e}(i,j,t)/(n_x \times n_y)$.
We gather that ${\rm <T_{He^{++}}>}$ reaches $2.4\times 10^7$ K
while ${\rm <T_{i}>}$ and ${\rm <T_{e}>}$ stay around
their initial average values of $1.2\times 10^7$ K.
Note at the domain edges initial temperature is $2\times 10^7$ K and
the temperature varies with y-coordinate according to Eq.\ref{eq2}. 
All panels show data for the Run 4, which has $n_x=5000$, $n_y=200$ and
$t_{\rm end} \omega_{\rm { cp}}=75$.

\begin{figure}
\includegraphics[width=\columnwidth]{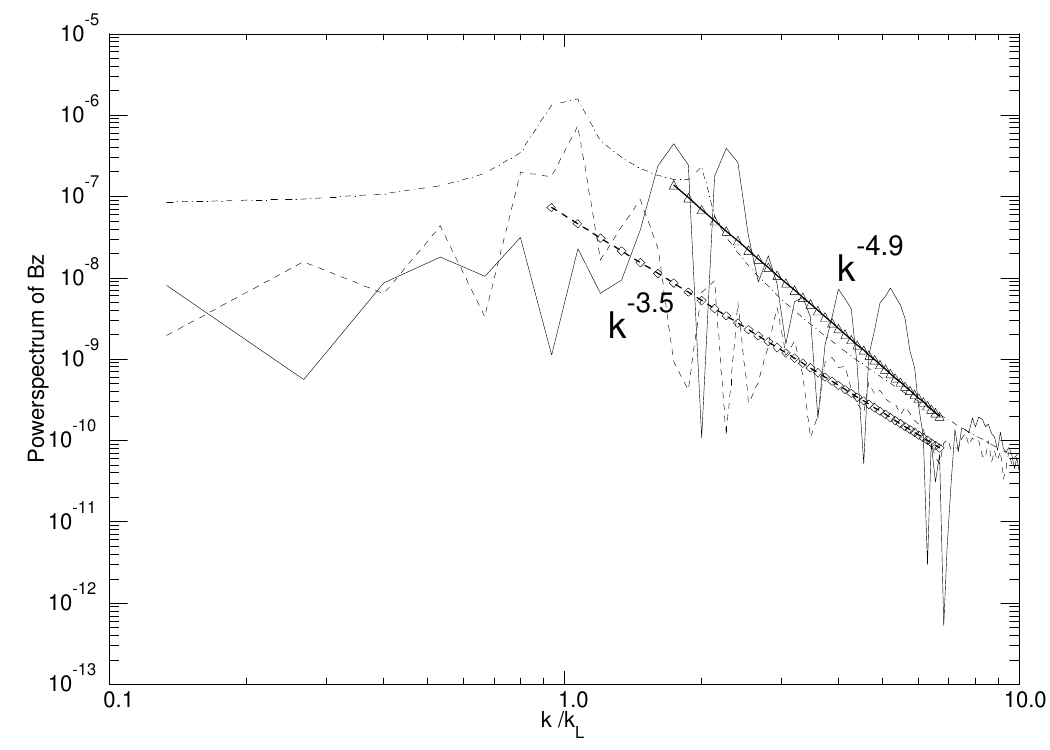}
    \caption{Scaling of the magnetic fluctuations power spectrum
with wavenumber $k$. We plot Fast Fourier Transform  of
$B_z(x,y=y_{\rm max}/2,t=t_{\rm max})$ (solid line);
$B_z(x,y=y_{\rm max}/20,t=t_{\rm max})$ (dashed line);
$B_{z,A}(x)$, according to Equation \ref{eq10},
 (dash-dotted line). Here data is from the Run 4.}
    \label{fig3}
\end{figure}
In Fig.\ref{fig3} we show scaling of the magnetic fluctuations power spectrum
with wavenumber $k$. In particular similar to \citet{ofman10}
we plot Fast Fourier Transform (FFT) of
$B_z(x,y=y_{\rm max}/2,t=t_{\rm max})$ (solid line);
$B_z(x,y=y_{\rm max}/20,t=t_{\rm max})$ (dashed line);
$B_{z,A}(x)$, according to Equation \ref{eq10}, (dash-dotted line).
Thus solid line shows $B_z$ fluctuations power spectrum
in the middle of the high-density region;
dash line shows the same but for a homogeneous, edge of the domain;
dash-dotted shows the same using the following analytical expression:
\begin{multline}
B_{z,A}(x)=\sum_{i=0}^{N-1}0.15 B_{0x} (i+1)^{-0.5}  F \\
\sin\left(\left[k_{\rm L}+
\frac{(2 k_{\rm L}-k_{\rm L})(i)}{N-1}\right] x
+0.05 \pi R(i)\right),
\label{eq10}
\end{multline}
where $k_L$ is left hand polarized wave number
calculated for $\omega_d=0.3 \omega_{ cp}$, using the general expression
 $$k_L=(\omega_{d}/c)\left(1.0 + (\omega_{pe}^2+\omega_{\rm { pp}}^2)/((\omega_{ce}+\omega_{d})(\omega_{ cp}-\omega_{d}))  \right)^{1/2}
 $$
with 
$\omega_{d}=0.3\omega_{ \rm cp}$.
{ We note that the above expression of $k_L$ is based on the dispersion 
relation for L- and R- polarized DAWs, which are as following \citep{dt11}
$$
k=\frac{\omega}{c}\left(1+\frac{\omega_{pe}^2+\omega_{pp}^2}{(\omega_{ce}\pm \omega)
(\omega_{cp} \mp \omega)}\right)^{1/2},
$$
here upper signs are for the L-polarization and lower signs for the R-polarization.
The latter equation indicates 
that for L-polarization possible physical
 frequency range is $0<\omega<\omega_{cp}$ and that there is
a ion-cyclotron resonance at $\omega=\omega_{cp}$.
Protons can resonate because they rotate in the same direction as
the electric field vector of the wave. 
In the case of 
R-polarization (not considered here) possible physical range is $0<\omega<\omega_{ce}$ with an electron-cyclotron 
resonance present at $\omega=\omega_{ce}$. }

In Fig.\ref{fig3} $B_{z,A}(x)$ with dash-dotted line
shows two expected peaks around $\omega_d=0.3 \omega_{ cp}$
(at location $k=k_L$)
and $\omega_d=0.6 \omega_{ cp}$ (at location $k=2 k_L$).
We also see the steepening of 
magnetic fluctuations power spectrum
in the higher-density regions, possibly due to wave refraction, as in previous Hybrid simulation results \citep{ofman10}, but now PIC runs produce much steeper slopes than in Hybrid runs. For example, in 
Fig.\ref{fig3} we see the scaling as $k^{-3.5}$ and $k^{-4.9}$
for the in low density, $B_z(x,y=y_{\rm max}/20,t=t_{\rm max})$,
and high density, $B_z(x,y=y_{\rm max}/2,t=t_{\rm max})$,
regions, respectively. Corresponding power-law indexes for the 
\citep{ofman10} Hybrid run (see their Fig.15) are $-1.7$ and $-2.5$.
Thus we see additional steepening and that
electron-scale physics, which can only be resolved
by a PIC simulation, has a notable effect of DAW spectrum evolution.
{ As mentioned earlier, this comparison should taken with care
because the considered physical parameters in the two studies are
different: solar wind and flaring coronal loop tops.} 

\begin{figure*}
\begin{center}
  \makebox[\textwidth]{\includegraphics[width=1.745\columnwidth]{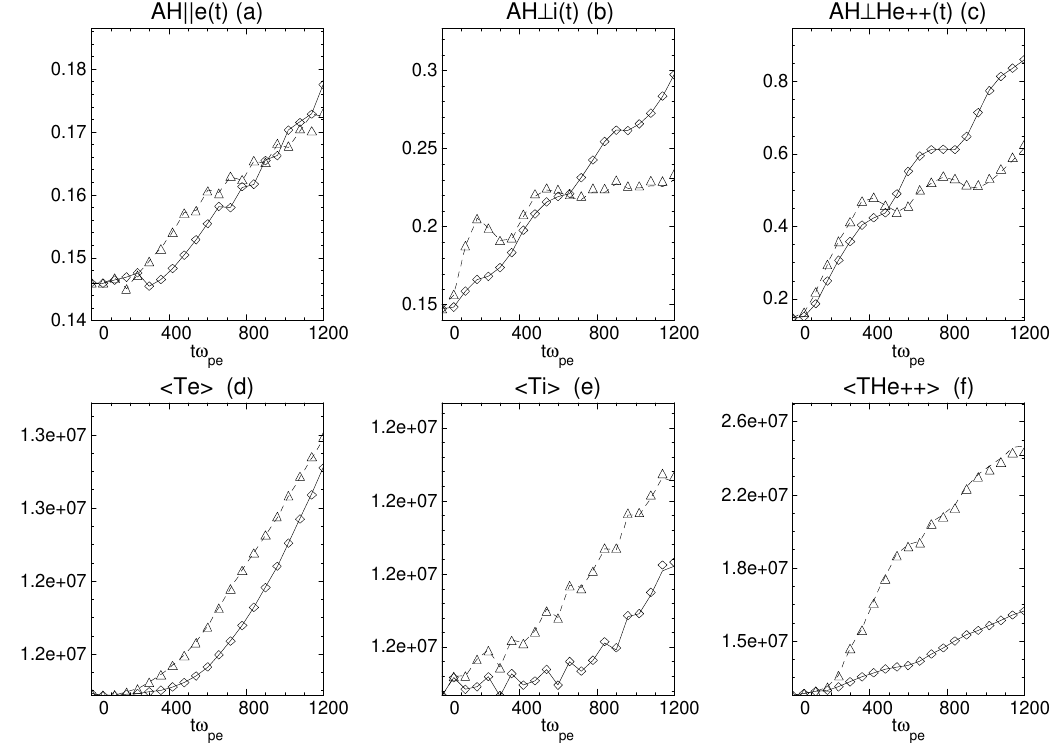}}
\end{center}
    \caption{In panel (a) we plot the following quantities
$AH_{\parallel,e}(t)$ for Run 1 (solid line);
$AH_{\parallel,e}(t)$ for Run 2 (diamonds);
$AH_{\parallel,e}(t)$ for Run 3 (dashed line);
$AH_{\parallel,e}(t)$ for Run 4 (triangles).
Panels (b),(c),(d),(e),(f) are follow the same pattern,
but are for $AH_{\perp, i}(t)$, $AH_{\perp, He}(t)$,
$<T_e(t)>$, $<T_i(t)>$, $<T_{He^{++}}(t)>$, respectively.}
    \label{fig4}
\end{figure*}
In Fig.\ref{fig4} we put to test 
 Eq.(\ref{eq7}). In particular we would like
to verify that Eq.(\ref{eq7}) inputs the same power via
DAW driving as a single harmonic case considered in
\citet{dt11}.
In panel (a) we plot the following quantities
$AH_{\parallel,e}(t)$ for Run 1 (solid line);
$AH_{\parallel,e}(t)$ for Run 2 (diamonds);
$AH_{\parallel,e}(t)$ for Run 3 (dashed line);
$AH_{\parallel,e}(t)$ for Run 4 (triangles).
We make two observations from panel (a) that:
(i) $AH_{\parallel,e}(t)$ for Run 1 and Run 2
overlap to a plotting precision, which means
that single harmonic and   a sum of 512 single
harmonics with amplitudes 512 times smaller than Run 1 produces identical
results;
(ii) $AH_{\parallel,e}(t)$ for Run 3 and Run 4
overlap to a plotting precision, which means
using 128 harmonics as per Eqs.\ref{eq5}-\ref{eq6}
with $F$ from Eq.\ref{eq7}, i.e. changing number of harmonic
still delivers the same power compared to 
Run 4 (512 harmonics).
A more general conclusion from panel (a) is that
as far is electron acceleration along the magnetic field
is concerned, the fraction of super-thermal
electrons remains broadly similar
with one uses single, 128, and/or 512 harmonics
with frequencies in the considered range.
Broadly similar conclusion can be reached 
about average temperature $<T_e(t)>$ dynamics in panel (d).
In other words broadband spectrum of DAWs does not affect much 
$AH_{\parallel,e}(t)$ dynamics.
Situation is quite different when we look at
dynamics of the proton and ${\rm He^{++}}$ hearing across the
magnetic field indexes $AH_{\perp, i}(t)$ and
$AH_{\perp, He}(t)$ shown in panels (b) and (c)
respectively.
Although as in panel (a) we see that Run 1 (solid line) and 
Run 2 (diamonds) data
as well as Run 3 (dashed line) and Run 4 (triangles) 
data overlap to a plotting precision,
Run 1/Run 2 attain higher values of $AH_{\perp, i}(t)$ and
$AH_{\perp, He}(t)$ at later times
compared to Run 3/Run 4.
As these indexes capture fraction of accelerated particles,
this result supports a conclusion that single harmonic (Run 1) or 128 harmonics
with 128 smaller amplitude (Run 2)
are more efficient in accelerating { protons} and ${\rm He^{++}}$, compared
to broadband spectrum of waves case (Run 3 and Run 4).
This can be understood by the fact that in 
Run 1/Run 2 DAW act coherently, while in Run 3/Run 4
there is also effect of 5\% random phase present which "washes out"
efficiency of particle acceleration which seems to be a coherent process
based on Landau resonance.
{ The heating due to DAWs is normally non-resonant, 
while the heating due to left-hand polarized ion-cyclotron waves is 
resonant with the resonant condition affected by the Doppler shift. 
We note that in our case Doppler shift is zero, as no drifts/flows of
particles present.}
In panels (e) and (f) we see reverse situation compared to
panels (b) and (c) in that broadband Run 3/Run 4 attain higher average
temperatures compared to coherent Run 1/Run 2.
This then clearly indicates that broadband spectrum that contains
cyclotron resonance produces profuse heating of ${\rm He^{++}}$ and to some
extent protons.
As summary of Figure \ref{fig4} we remark that
the frequency spectrum case does not seem to affect electron acceleration fraction in the like-to-like cases, but few times larger percentage of ${\rm He^{++}}$ heating is seen due to ion cyclotron resonance;

\begin{figure*}
\begin{center}
  \makebox[\textwidth]{\includegraphics[width=1.745\columnwidth]{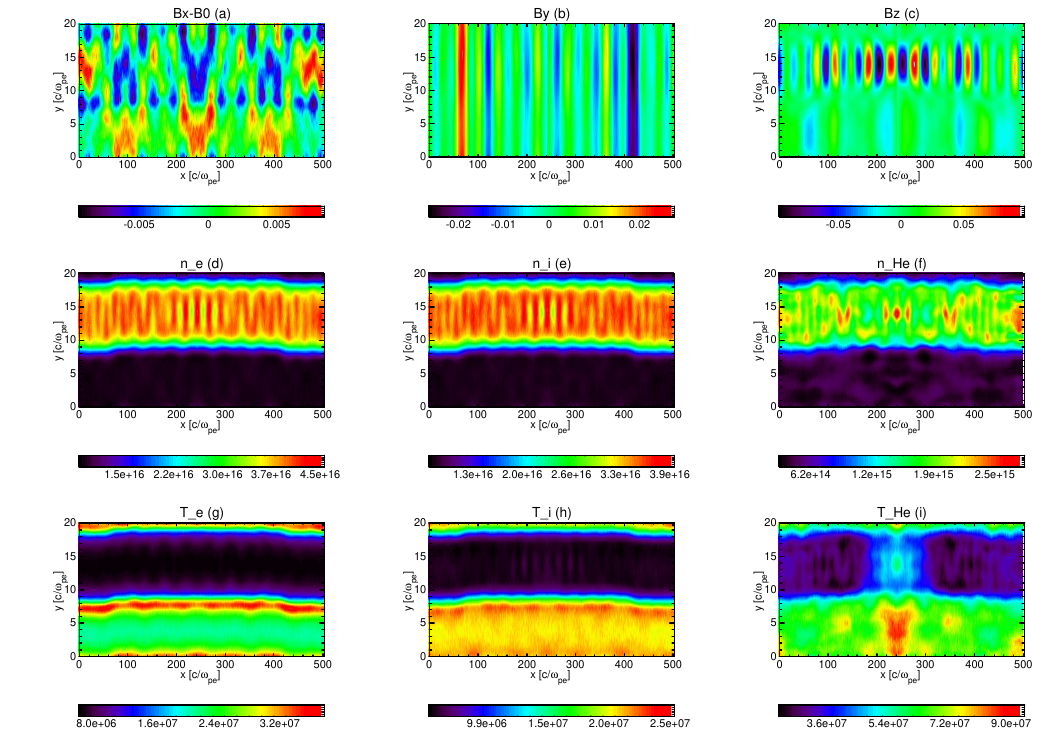}}
\end{center}
    \caption{Similar to Figure \ref{fig1} but for Run 5.}
    \label{fig5}
\end{figure*}
In Fig.\ref{fig5} we show panels similar to Figure \ref{fig1} but for Run 5.
Purpose of Run 5 is to study how multiple DAW front collisions
will affect particle acceleration. Thus in Run 5 we set
$t_{\rm end} \omega_{\rm { cp}}=750$ with 10-times longer simulation time, 
which insures that
the DAWs will collide at least 
9 times.
We note in panel (a) that perturbation of $B_x$ is no longer confined
the density gradient regions and spread across y-coordinate,
hence indicating oblique propagation across the magnetic field at small angle.
We note that panel (b) is similar to panel (b) in Figure \ref{fig1}.
The notable diffidence can be seen in panels (c)--(i) in that:
(i)
the whole density/temperature inhomogeneity has shifted its location
from $y=y_{\rm max}/2$ to larger values of y-coordinate;
(ii) noticeable bending of the entire 
background density/temperature inhomogeneity
structure can be seen, i.e. edges at $x=0$ and $x=x_{\rm max}$ have lower y-coordinates than $x=x_{\rm max}/2$, i.e. we see
development of kink oscillations when driving DAWs collide.
When seeing animated version of Fig.\ref{fig5} (not shown here)
the background density/temperature inhomogeneity
structure's $x=0$ and $x=x_{\rm max}$ edges move up and down in concert,
while the middle part $x=x_{\rm max}/2$ oscillates out of phase, giving
rise to kink-like motion.
This kinking seems to come 
from DAW wave collision, which seems reasonable and intuitive
from the first physical principles, and yet new according 
to the author's opinion.
In panels (g) and (h) we see significant increase of temperatures
in the density gradient regions, while for panel (i)
the temperature increase is near the DAW injection location at
$x=x_{\rm max}/2$,
we think, this is due to the 
ion cyclotron resonance for ${\rm He^{++}}$.

\begin{figure*}
\begin{center}
  \makebox[\textwidth]{\includegraphics[width=1.745\columnwidth]{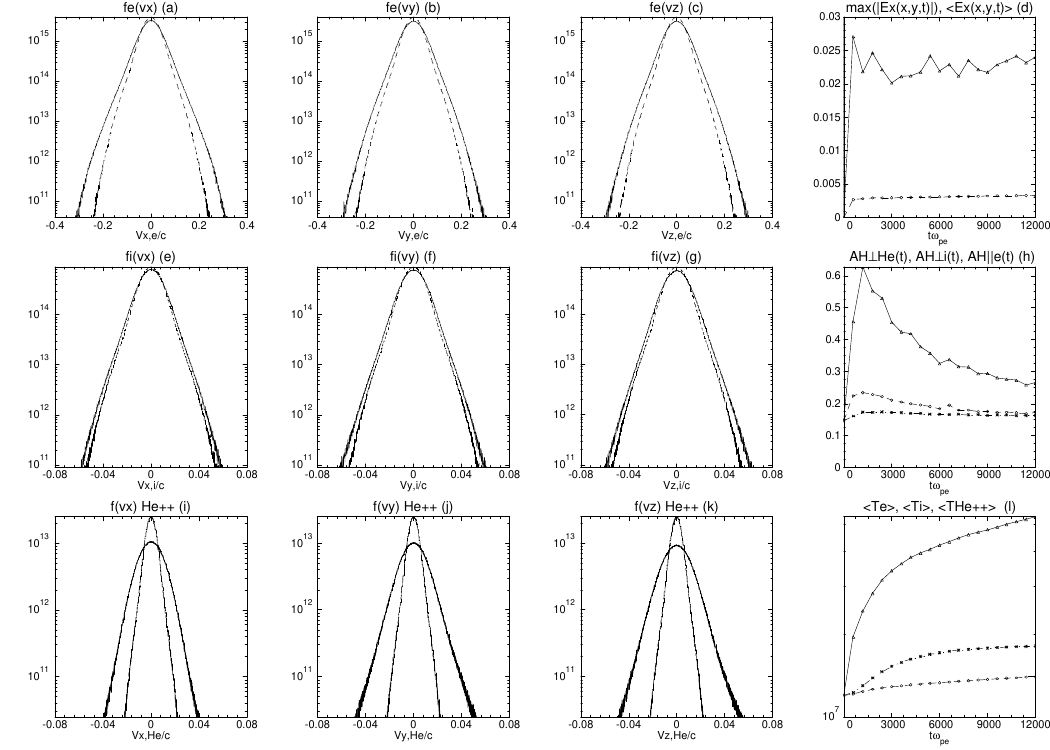}}
\end{center}
    \caption{Similar to Figure \ref{fig2} but for Run 5.}
    \label{fig6}
\end{figure*}
In Fig.\ref{fig6} we plot panels 
similar to Figure \ref{fig2} but for Run 5.
Commenting on the differences compared to Figure \ref{fig2} we note the
following new features:
(i) now electron distribution function is also broadened
in y- (panel (b)) and z- (panel (c)) directions. 
This is because of 10 times longer
driving by DAWs injects much larger amount of energy
into the system. Thus, electron population is heated up and we see
more isotropic velocity distributions in all three spatial directions;
(ii) we note in panel (h) an intermittent in time 
 electron, proton and ${\rm He^{++}}$ acceleration fractions. This is because intensive heating (plasma temperature increase) makes the-above-thermal-fraction smaller;
 Broadly speaking, when in Figure \ref{fig2} we see
 $AH_{\perp, He}(t)$ attains value of 0.63, i.e.
63\% of ${\rm He^{++}}$ become super-thermal.
But in Figure \ref{fig6} $AH_{\perp, He}(t)$ 
after attaining is peak value it starts to decease
down to 0.25,  this is because plasma temperature increase makes the-above-thermal-fraction (i.e denominator increase) smaller;
(iii) now much higher temperatures are attained (panel (i))
by all plasma species namely
$<T_e>=1.8\times 10^7$ K, $<T_i>=1.4\times 10^7$ K, $<T_{He^{++}}>=5.3\times 10^7$ K.
The temperature increase for 
${\rm He^{++}}$ is the highest of all three species, 
because of ion-cyclotron resonance condition is met.
In  this context, we mention a new study  that
 analyses relativistic electron measurements obtained by the High Energy Telescope (HET) aboard Solar Orbiter in
the energy range from 200 keV to 10 MeV \citep{fleth23}.
The latter authors compile a list of 21 greater 200 keV energy
 electron time periods
of enhanced flux. Out of this list  they detect
 3 events with clearly anisotropic electron pitch-angle distributions 
 above 1 MeV. Also 5 other events have been seen within 0.5 AU  
 from the Sun that show no discernible anisotropy. 
Hence, upcoming data from HET aboard Solar Orbiter will be able to
test predictions made from numerical simulation studies such as this one.
In our case in Figure \ref{fig1} electron distribution function 
is anisotropic, while in Figure \ref{fig6} it is isotropic for the reasons explained above.

\begin{figure}
\includegraphics[width=\columnwidth]{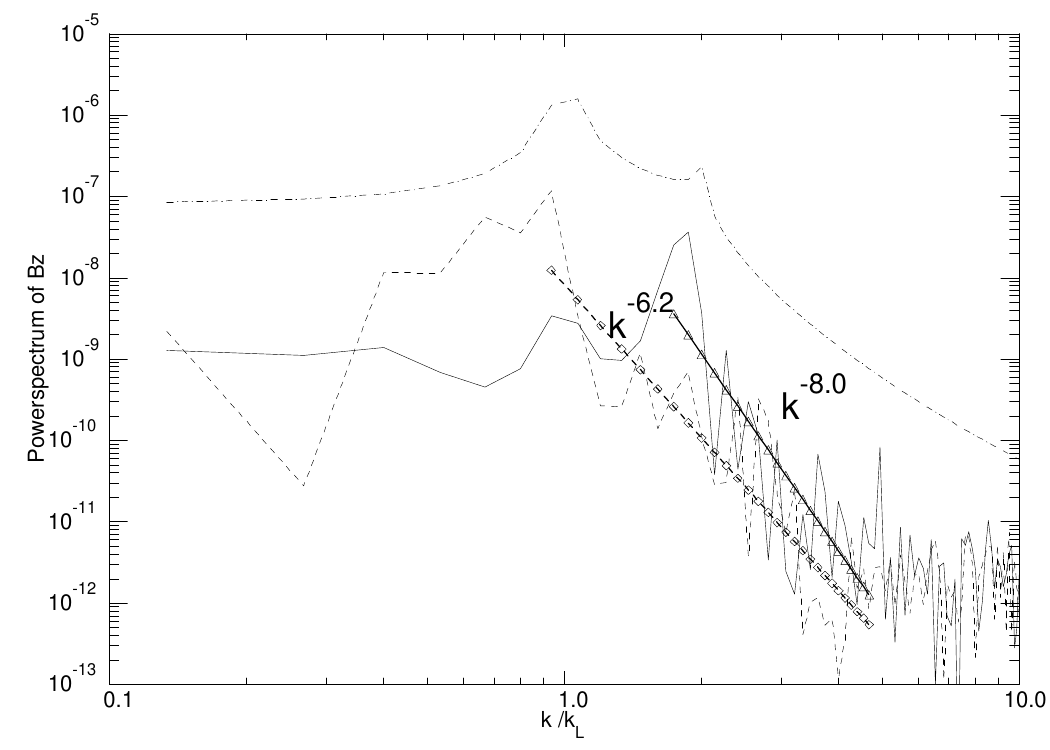}
    \caption{Similar to Figure \ref{fig3} but for Run 5.}
    \label{fig7}
\end{figure}
In Fig.\ref{fig7} we plot 
similar physical quantities  to Figure \ref{fig3} but for Run 5.
We note that now we see
{\it further} steepening of 
magnetic fluctuations power spectrum  both
in the higher-density regions and the domain edges.
In particular, Fig.\ref{fig7} we see the scaling as $k^{-6.2}$ and $k^{-8.0}$
for the in low density, $B_z(x,y=y_{\rm max}/20,t=t_{\rm max})$,
and high density, $B_z(x,y=y_{\rm max}/2,t=t_{\rm max})$,
regions, respectively.
It is difficult to comment on the specific reason for the spectrum slope
steepening as a result of multiple DAW collisions, but generally
steepening of the slope means increased 
dissipation. Hence steepened slopes means
that multiple DAW collisions seeded turbulence which
made dissipation more intense. 
We admit, this is a speculative argument based on an intuitive judgement.

\begin{figure*}
\begin{center}
  \makebox[\textwidth]{\includegraphics[width=1.745\columnwidth]{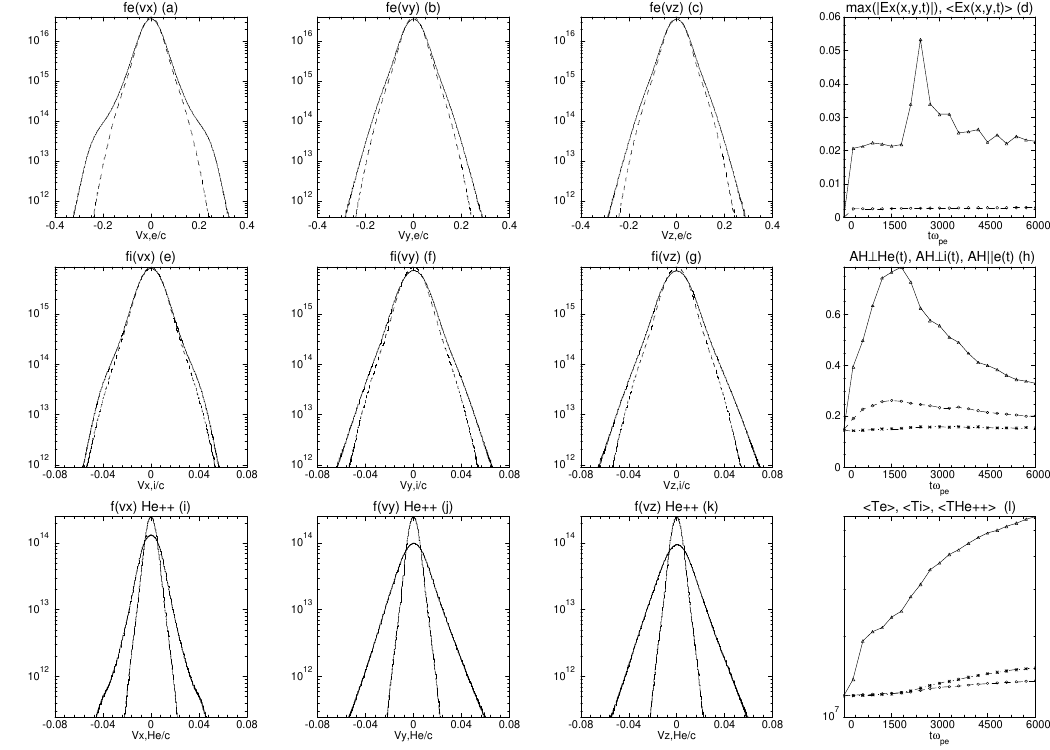}}
\end{center}
    \caption{Similar to Figure \ref{fig2} but for Run 6.}
    \label{fig8}
\end{figure*}
In Fig.\ref{fig8} we show panels  Figure \ref{fig2} but for Run 6,
which has 
the length-scale of the density gradient in $y$-direction increased
10 times to approximately "ion"-scales
$d n/ dy\simeq 40 c/\omega_{\rm { pe}}$. 
Hence in Run 6 there are
$x=5000$ and $y=2000$ grid points in each direction.
The main purpose of this run is to study how particle acceleration is
modified by up-scaling the transverse density gradient spatial 
size 10 times.
There are two notable differences compared to Run 4 (Figure \ref{fig2})
in that (i) 
now ${\rm He^{++}}$ peak heating by ion-cyclotron 
is higher with the maximum of 
$AH_{\perp,i}(t)$ now attaining value of 0.79 (79\%);
(ii) The maximal value of $<T_{He^{++}}>=5.6\times 10^7$ K, which is also a higher value.
From these finding we conclude that increasing density gradient 10 times
enhances ion cyclotron heating for ${\rm He^{++}}$ ions which is evidenced by
the higher super-thermal percentages and also the higher 
the temperature gained
by the ${\rm He^{++}}$ plasma species.
Hence the main conclusion of this Run 6 is that
increasing density gradient scale
across the magnetic field 10 times 
(almost to ion scales) does not affect the particle acceleration
features by DAWs. We stress this point because the usual
criticism  
such numerical simulations attract (Private Communication with
various Plasma Physics colleagues) is that
when one up-scales density gradient -- the effect of
parallel electric field generation and particle acceleration
by DAWs goes away. Here we demonstrate otherwise.

\begin{figure*}
\begin{center}
  \makebox[\textwidth]{\includegraphics[width=1.745\columnwidth]{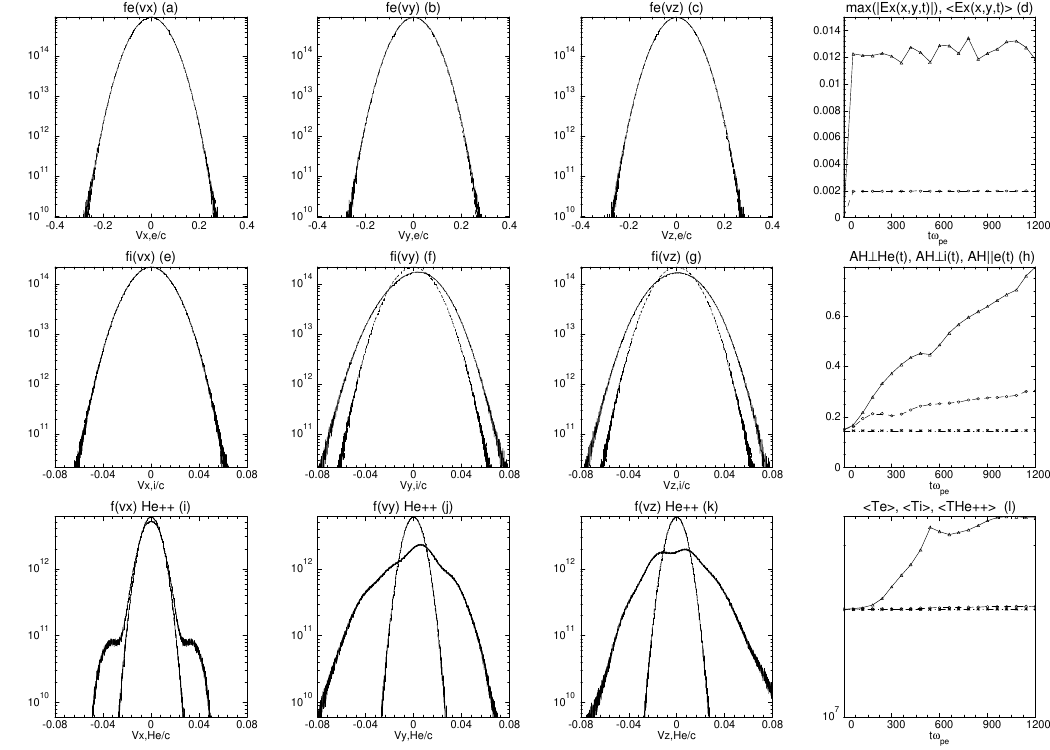}}
\end{center}
    \caption{Similar to Figure \ref{fig2} but for Run 7.}
    \label{fig9}
\end{figure*}
In Fig.\ref{fig9} we show  
similar physical quantities  to Figure \ref{fig2} but for Run 7.
We note the following differences compared to
Run 4 (Fig.\ref{fig2}) in that 
in this homogeneous plasma case, i.e. no
transverse density/Alfven speed inhomogeneity
across the magnetic field case, there is no 
electron acceleration present at all 
by the end simulation time, which
supports the conclusion that the electron acceleration by DAWs
is purely caused by the transverse plasma density
inhomogeneity (panel (a)).
There is an indication of 
(i) ${\rm He^{++}}$ beams being formed
along the magnetic field
(panel (i)); and
(ii) the bulk heating still occurring across the magnetic field 
due to cyclotron resonance in  panels (j) and (k),
by the end simulation time. 
The former seems not well understood
from physical grounds as to why homogeneity should support
ion-cyclotron magnetic field-aligned resonant beams, while
the latter is well understood by the said resonance.

\begin{figure*}
\begin{center}
  \makebox[\textwidth]{\includegraphics[width=1.745\columnwidth]{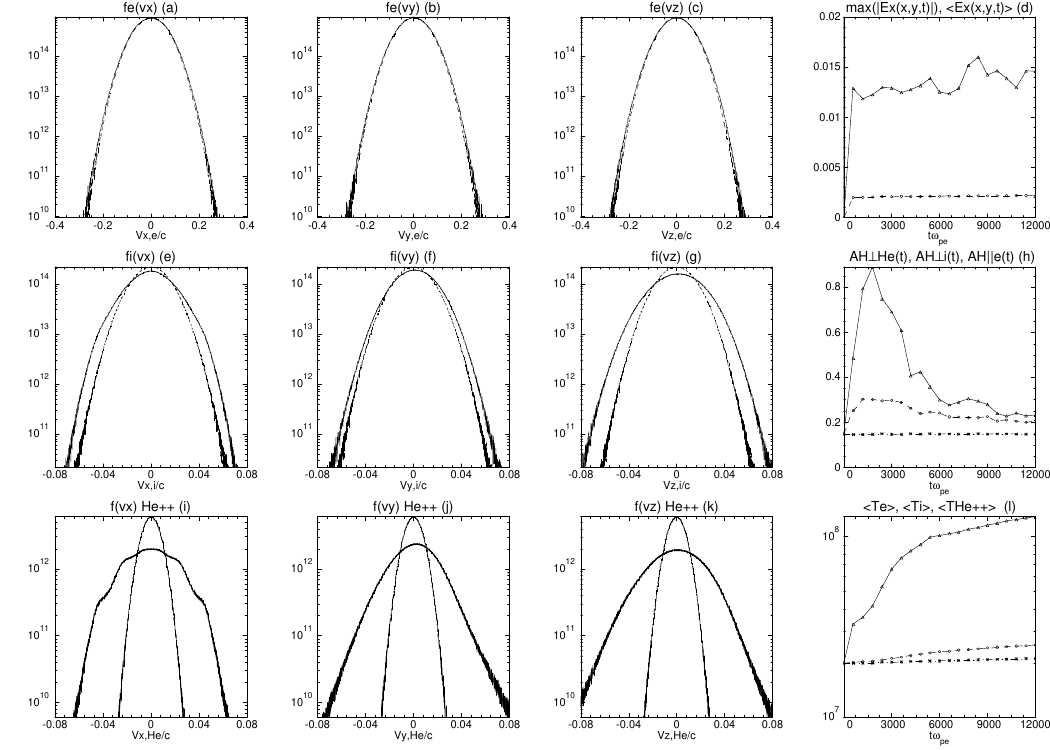}}
\end{center}
    \caption{Similar to Figure \ref{fig2} but for Run 8.}
    \label{fig10}
\end{figure*}
In Fig.\ref{fig10} we show
similar physical quantities  to Figure \ref{fig2} but for Run 8.
We note the following differences compared to
Run 5 (cf. Fig.\ref{fig6}) in that
in this homogeneous plasma case,
despite 10 times longer end-simulation time of 
$t_{\rm end} \omega_{\rm { cp}}=750$
we still not see any B-field aligned
electron acceleration (panel (a)).
There is an indication of ${\rm He^{++}}$ distribution functions
become isotropic in $V_x$, $V_{ y}$, $V_z$ directions
(panels (i),(j),(k)).
From panel (h) we deduce that
$AH_{\perp, He}(t)$ attains value of 0.89, i.e.
89\% of ${\rm He^{++}}$ become super-thermal.
From panel (l) we note that average temperatures of all
plasma species became increased:
$<T_e>=2.1\times 10^7$ K, $<T_i>=2.5\times 10^7$ K, $<T_{He^{++}}>=1.3\times 10^8$ K.
Especially large increase is seen in $<T_{He^{++}}>$ due to continuous 
action of ion-cyclotron resonance.

\begin{figure*}
\begin{center}
  \makebox[\textwidth]{\includegraphics[width=1.745\columnwidth]{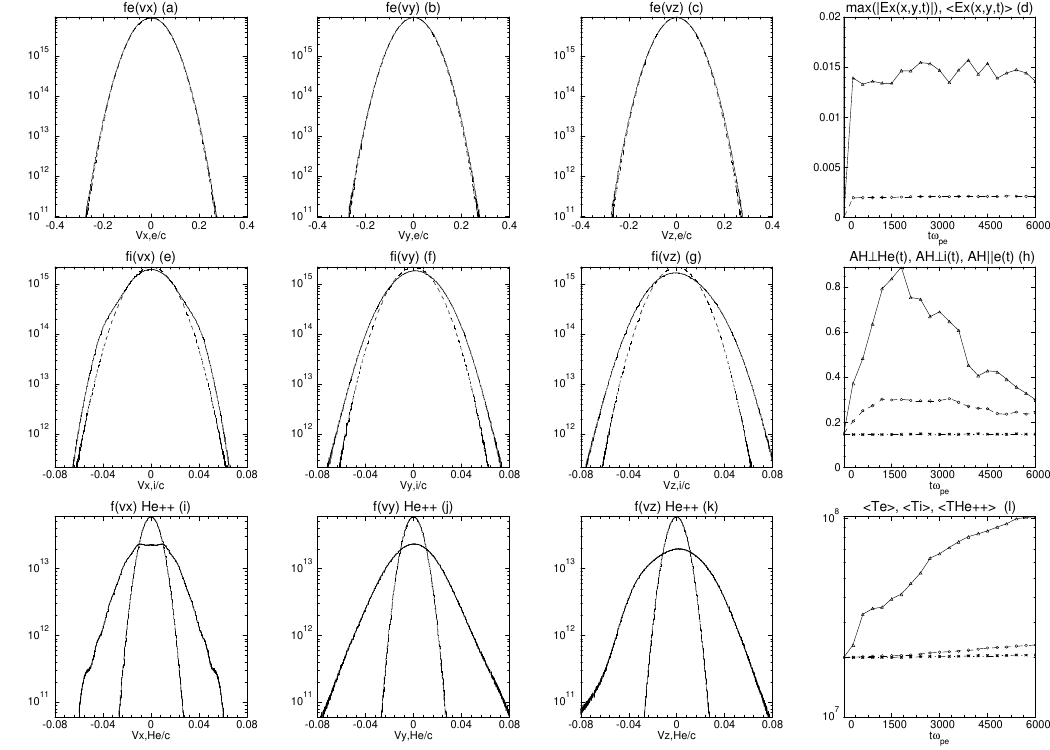}}
\end{center}
    \caption{Similar to Figure \ref{fig2} but for Run 9.}
    \label{fig11}
\end{figure*}
In Fig.\ref{fig11} we show
similar physical quantities  to Figure \ref{fig2} but for Run 9.
We note the following differences compared to
Run 6 (cf. Fig.\ref{fig8}) in that
in this homogeneous plasma case,
despite 10 times larger domain-size, hence
10 times larger density gradient in transverse to the magnetic
field y-direction, and $t_{\rm end} \omega_{\rm { cp}}=375$
(maximum time that was computationally feasible), we again
note no electron acceleration along x-direction (panel (a)).
The other conclusions that we reached for 
Fig.\ref{fig10} (Run 7) are applicable here in this Run 9 too.
Namely that $AH_{\perp, He}(t)$ is peaking at 0.89; and
isotropization of ${\rm He^{++}}$ distribution functions
in $V_x$, $V_{ y}$, $V_z$ directions is seen;
as well as
average temperatures of all
plasma species attaining values of 
$<T_e>=2.05\times 10^7$ K, $<T_i>=2.3\times 10^7$ K,
 $<T_{He^{++}}>=1.01\times 10^8$ K.

\section{Conclusions}

{ { We would like to comment on the feasibility} of the DAWs driving
considered in this work and more generally what is known what { the}
type of waves solar and stellar flares can generate.
In the solar flare context, the first study which initiated 
investigation of electron acceleration by a 
single harmonic with sub-proton cyclotron DAWs 
$\omega_d=0.3\omega_{\rm cp}$ (to avoid proton cyclotron resonance) 
 in the transversely inhomogeneous
plasma is by \citet{2005A&A...435.1105T}.
Subsequently, \citet{Fletcher_2008} explored a possibility when flare-generated 
reconfiguring coronal field launches a torsional AW pulse.
We refer { the} reader to Figure 1 from \citet{Fletcher_2008}
which shows that AWs are launched at the loop top
and propagate towards the foot-points. They study in detail
feasibility of electron acceleration by estimating
various parameters such as parallel electric field based on
two-fluid theory. Although \citet{Fletcher_2008} refer to AW pulse,
estimates are made based on plasma kinetic dispersion
relations, which are based on harmonic waves, like the ones
considered in the present work. We believe initial AW pulse, 
rapidly phase-mixes on the
transverse inhomogeneity and reaches short, kinetic scales.
As can be seen from Fig.\ref{fig1}, panels (b) and (c), 
by the end simulation time DAWs
develop only 3 wavelength by the end simulation time.
So it is plausible that inital AW pulse generates L-hand polarized
DAWs on such time-scale. Also, it is known that in the sub-proton
cyclotron frequency regime, linearly polarized AWs 
naturally decay into and L- and
R-hand circularly polarized DAWs \citep{kt86} because their phase speeds
are different.

\citet{hm16} suggest that electromagnetic ion cyclotron (EMIC) waves are important for electron acceleration and loss from the Earth
radiation belts. They suggest that these waves are generated by unstable ion distributions that form during geomagnetically disturbed times. 
\citet{hm16} show that magnetosonic waves could be a source of EMIC waves as a result of propagation and a process of linear mode conversion.
It is feasible the same mechanism also works in solar corona,
with solar flares playing the same role as geomagnetic storms,
which is essentially magnetic reconnection.
{ The observed MHD wave amplitudes  of $0.05 B_0$, 
i.e. 5\% of the background magnetic field are typically observed. }
Such waves are routinely observed in the solar corona 
and also found in MHD simulations e.g. \citet{li22} 
reported the evolution of a traveling kink pulse to a 
standing kink wave along coronal loops that has been induced by a solar flare.
As per their Figure 5 amplitude can be estimated
roughly as zero-to-maximal deflection divided by 2 minutes (120 s)
$V=\Delta x / \Delta t = 12 \; {\rm Mm} / 120 \; {\rm s}=100 \; {\rm km/s}$
which is about 10 \% of their estimated Alfven seed of 950 km/s.
{Above we discussed 
MHD waves that were observed on spatial scales of 10,000-100,000 km, 
while the wavelengths of the waves considered in the present study are on order of $10^{-2}$ m, 8-10 orders of magnitude shorter wavelength than the MHD waves. It may seem that the discussion of MHD scale waves is not relevant here,
but presently only such scale waves are observed because DAWs are too small scale to be observable presently.
Here we refer the reader back to the relevant discussion of
unrealsitically high wave amplitudes in the kinetic regime considered in this study, 
and the objective computational reasons for that, stated
in a paragraph just above equation \ref{eq5}.
 Moreover, there are some other mechanisms that can effectively excite small-scale KAWs in non-uniform media, such as the resonant mode conversion of AWs into KAWs \citep{Xiang_2019} and the density striation instability 
 \citep{Wu_2013,chen_2015}. Therefore, it is meaningless to argue about the magnitude of the amplitude of the unmeasurable KAWs in the solar corona.}

\citet{Kryshtal2020} considered generation of low-frequency plasma waves in the
lower/middle chromosphere region of magnetic 
loop foot-points when in the pre-flare state plasma.
They showed that the generation of KAWs and kinetic ion-acoustic 
waves can occur both, in plasma with Coulomb conductivity and in the 
presence of small-scale Bernstein turbulence in the solar chromospheric setting. \citet{Ofman_2022}
model the excitation and dissipation of slow magneto-sonic MHD
waves in hot coronal flaring loops according to EUV observations 
using a 3D magnetohydrodynamic (MHD) visco-resistive simulations.
\citet{PROVORNIKOVA2018645} contributes to an understanding of
 physical
properties of observed solar coronal 
perturbations by investigating the excitation of 
waves by hot plasma injections from below and the evolution of
flows and wave propagation along the solar coronal loop. 
Such ejections are can presumably come from solar flares.
We remark that DAWs are observed in situ routinely in 
Earth magnetosphere \citep{Lysak2023}
 but the Sun is too far away for DAWs  to be detected remotely. 
The 1/f spectrum of AWs naturally appears due to turbulent 
scale cascade and is a sign of scale-free (fractal-like) behaviour. 

We would like to comment about feasibility of the spatial 
scale of transverse inhomogeneity considered in this work, namely,
$4-40 c/\omega_{\rm { pe}}$. In SI units for the
physical parameters considered this corresponds to  
$0.21-2.1$ m. For 
the real proton (for the mass ratio
$m_p/m_e=1836$) and electron inertial length are
$c/\omega_{\rm pp}=2.3$ m and $c/\omega_{\rm pe}=0.05$ m, respectively.
So in fact when we state that we consider inhomogeneity
on "ion scale" $40 c/\omega_{\rm { \rm pe}} =2.1$ m, it
is not very different from the realistic $c/\omega_{\rm pp}=2.3$ m. 
However, neither Parker Solar Probe nor Solar Dynamics
Observatory can resolve magnetic loop transverse threads of
such small width. The reason we consider such small scales
is because current PIC models cannot consider realistic
physical scales. This does not mean that such
fine-scale transverse structuring does not exist and
future observations may well detect it.}

In this work we extended our previous results
 \citep{2005A&A...435.1105T,dt11,dt12}) in five-ways:
(i) instead of exciting DAWs at the left edge of the domain which mimics
loop top we now drive DAWs in the middle of the domain, which
more realistically represents DAWs generated at the solar coronal loop top
during a flare;
(ii) we now consider wide spectrum of DAWs as specified in Table \ref{t1};
(iii) we added ion-cyclotron resonant ${\rm He^{++}}$ species;
(iv) we considered cases when DAW collide multiple times with now 
10 times longer 
in time  than our previous numerical runs;
(v) we considered cases when the density gradient scale, and 
commensurately  transverse
to the magnetic field domain spatial size, is now
increased 10 times compared to our previous results. 
Hence we studied cases when transverse density gradient is  in the range ${ 4-40} c/\omega_{\rm { pe}}$ and DAW driving frequency is $0.3-0.6\omega_{\rm { cp}}$. 
Detailed discussion of the results established in this work can be found in
the results section. Here we mention only the main findings as following: 
(i) The considered 
DAW driving frequency spectrum
 $0.3-0.6\omega_{\rm { cp}}$  does not 
 affect electron acceleration fraction in the like-to-like cases, 
 but few times larger percentage of ${\rm He^{++}}$ heating is seen due to ion cyclotron resonance; (ii) When we consider DAW multiple collisions, then much larger electron and ion acceleration fractions  are seen, but the process is 
 intermittent in time and more isotropic velocity distributions are also found.
  We attribute this to 
 intensive heating (plasma temperature increase) makes the-above-thermal-fraction smaller; 
 (iii) We witness formation  of kink oscillations when DAWs collide, which
 seems a novel effect;
 { We observe noticeable bending of the entire 
background density/temperature inhomogeneity
structure, i.e. edges at $x=0$ and $x=x_{\rm max}$ have lower y-coordinates than $x=x_{\rm max}/2$, and we see
development of kink oscillations when driving DAWs collide.
We see also that
the background density/temperature inhomogeneity
structure's $x=0$ and $x=x_{\rm max}$ edges move up and down in concert,
while the middle part $x=x_{\rm max}/2$ oscillates out of phase, giving
rise to a kink-like motion.
Due to the limitation of data storage, we could only output
20 snapshots of the kink oscillations from time zero to the final simulation
time. This is not sufficient to infer the period of the kink oscillation,
but we can confirm that this is the fundamental harmonic, with 
the maximal oscillation occurring at $x=x_{\rm max}/2$.}
(iv) We study scaling of the magnetic fluctuations power spectrum 
and find that steepening in the higher-density regions occurs, 
due to wave refraction. This means that inclusion of
electron-scale physics has a notable effect of DAW spectrum evolution.

\section*{Acknowledgements}

Calculations were performed using the Sulis Tier 2 HPC platform hosted by the Scientific Computing Research Technology Platform at the University of Warwick. Sulis is funded by EPSRC Grant EP/T022108/1 and the HPC Midlands+ consortium.
This research utilized Queen Mary's Apocrita HPC facility, supported by QMUL Research-IT.  \url{http://doi.org/10.5281/zenodo.438045}.

\section*{Data Availability}
All data used in this study has been generated by Particle-in-cell code for plasma physics simulations called 
EPOCH, cf. \citet{Arber:2015hc}, available for download from GitHub repository
\url{https://github.com/Warwick-Plasma/epoch}.
The derived data generated in this 
research will be shared on reasonable request to the corresponding
author.



\bibliographystyle{mnras}
\bibliography{paper84} 



\bsp	
\label{lastpage}
\end{document}